\documentclass{aa}  

\usepackage[switch]{lineno}
\usepackage{graphicx}
\usepackage{natbib}

\usepackage{supertabular}

\usepackage{txfonts}

\usepackage[]{hyperref}

\usepackage{lscape}

\usepackage[usenames]{color}

\newcommand{\bcep}{$\beta$~Cep }
\newcommand{\dsct}{$\delta$~Sct }

\usepackage{amssymb}
\usepackage{mathrsfs}

\begin{document} 


   \title{The CubeSpec space mission}
   
   \subtitle{I. Asteroseismology of massive stars from time series optical spectroscopy: science requirements and target list prioritisation}

   \titlerunning{CubeSpec: time-series spectroscopy for massive star asteroseismology}
   
   \author{D. M. Bowman\inst{1} 
          \and
          B. Vandenbussche \inst{1}
          \and
          H. Sana \inst{1}
          \and
          A. Tkachenko \inst{1}
          \and
          G. Raskin \inst{1}
          \and
          T. Delabie \inst{2}
          \and
          B. Vandoren \inst{2}
          \and
          P. Royer \inst{1}
          \and
          S.~Garcia \inst{1}
          \and
          T. Van Reeth \inst{1}
          \and
          the CubeSpec collaboration
          }

    \institute{Institute of Astronomy, KU Leuven, Celestijnenlaan 200D, B-3001 Leuven, Belgium \\
              \email{dominic.bowman@kuleuven.be} 
              \and
              Arcsec NV, Blijde Inkomststraat 22, B-3000 Leuven, Belgium
          }

   \date{Received 5 Oct 2021 / accepted 18 Nov 2021}

  
 
  \abstract
   {There is currently a niche for providing high-cadence, high resolution, time-series optical spectroscopy from space, which can be filled by using a low-cost cubesat mission. The Belgian-led ESA CubeSpec mission is specifically designed to provide space-based, low-cost spectroscopy with specific capabilities that can be optimised for a particular science need. Approved as an ESA in-orbit demonstrator, the CubeSpec satellite's primary science objective will focus on obtaining high-cadence, high resolution optical spectroscopic data to facilitate asteroseismology of pulsating massive stars.}
   {In this first paper, we aim to search for pulsating massive stars suitable for the CubeSpec mission, specifically \bcep stars, which typically require time series spectroscopy to identify the geometry of their pulsation modes.}
   {Based on the science requirements needed to enable asteroseismology of massive stars with the capabilities of CubeSpec's spectrograph, we combine a literature study for pulsation with the analysis of recent high-cadence time series TESS photometry to classify the variability for stars brighter than $V \leq 4$~mag and between O9 and B3 in spectral type. }
   {Among the 90 stars that meet our magnitude and spectral type requirements, we identify 23 promising \bcep stars with high-amplitude (non-)radial pulsation modes with frequencies below 7~d$^{-1}$. Using further constraints on projected rotational velocities, pulsation amplitudes and number of pulsation modes, we devise a prioritised target list for the CubeSpec mission according to its science requirements and the potential of the targets for asteroseismology. The full target catalogue further provide a modern TESS-based review of line profile and photometric variability properties among bright O9--B3 stars.}
   {}


   \maketitle


\section{Introduction}
\label{section: intro}

Massive stars are key astrophysical objects, as through their radiation and mechanical feedback they heat and enrich the interstellar medium and significantly contribute towards the dynamical and chemical evolution of their host galaxy \citep{Bromm2004c, deRossi2010d, Robertson_B_2010d}. Massive stars are the progenitors of stellar mass black holes and neutron stars. The coalescence of these exotic compact objects during the end phases of their evolution generate gravitational waves, which are now detectable and at the forefront of astronomical research thanks to the LIGO and VIRGO observatories \citep{Abbott_B_2016d, Abbott_B_2019m}. Despite their importance, the pre-supernova evolution of massive stars remains poorly constrained, with the some of the largest uncertainties related to the internal structure and mixing processes \citep{Langer2012}. 

The internal structure beneath the opaque surface of a single star is difficult to infer using traditional techniques in astronomy such as photometry or spectroscopy. Specifically, the rotation and mixing profiles, chemical diffusion, and convection physics within massive star interiors represent large theoretical uncertainties in current models \citep{Maeder_rotation_BOOK, Kippenhahn_BOOK, Meynet_rotation_BOOK}. Additional uncertainties in stellar evolution models include binarity \citep{Sana2012b, Kobulnicky2014}, rotation, the impact of winds and mass loss \citep{Ekstrom2012a, Georgy2013c, Vink2021b*}, and the amount and shape of the mixing in the core-boundary region during the main sequence phase of evolution \citep{Aerts2015a, Bowman2020c}. The latter is particularly important and is currently represented in modern stellar evolution codes by ad hoc prescriptions with specific free parameters (see e.g. \citealt{Martins2013c} for a comparison of different evolution codes). For massive stars, the lack of direct observational constraints produce fractional uncertainties for main-sequence lifetimes and helium core masses as much as 50\,\% \citep{Bowman2020c, Johnston2021b*}, which influences the nature of the compact object left behind \citep{Langer2012, Farrell2020a}. 

An excellent astronomical technique to directly measure the interior properties of stars, such as core masses, rotation and convective boundary mixing, is asteroseismology \citep{ASTERO_BOOK}. The eigenmode frequencies of a star are set by its structure, such that the forward modelling of observed pulsation mode frequencies allows one to constrain its interior physics \citep{Aerts2021a}. In asteroseismology, there are two main types of observational techniques to obtain a star's pulsation mode frequencies. Both involve obtaining a time series of observed quantities, from which Fourier analysis yields the pulsation mode frequencies. The first technique is time series photometry (i.e. a light curve), and the second is time series spectroscopy to measure line profile variability (LPV) of a star's spectral lines \citep{ASTERO_BOOK}. During its pulsation cycle, a star contracts and expands changing its radius and surface temperature. This generates periodic brightness changes in the photosphere which are seen as variability in a light curve. The complementary technique of time series spectroscopy allows one to measure the velocity fields produced during the pulsation cycle through the LPV of spectral lines in the line forming region. After the pulsation mode frequencies have been extracted from the time series data, they need to be identified in terms of spherical geometry to yield the unique radial order, $n$, angular degree, $\ell$, and azimuthal order, $m$. Subsequent modelling of the identified pulsation mode frequencies through quantitative comparison to a grid of theoretical frequencies yields constraints on a star's interior properties (see e.g. \citealt{Briquet2012, Aerts2019a}).

In the last decade, forward asteroseismic modelling of pulsation frequencies has strongly relied on time series photometry from space missions such as CoRoT \citep{Auvergne2009}, Kepler/K2 \citep{Borucki2010, Koch2010, Howell2014}, BRITE \citep{Weiss2014, Weiss2021a}, and more recently TESS \citep{Ricker2015}. The revolution in space photometry and abundance in high-precision, high-cadence and long-duration light curves has allowed asteroseismic studies to extract the interior properties of stars with masses between 1 and 25~M$_{\odot}$ using asteroseismic modelling \citep{Chaplin2013c, Bowman2020c, Aerts2021a}. Highlights include high-radial order gravity (g) modes used to determine precise (core) masses, ages and interior mixing profiles in slowly pulsating B (SPB) stars \citep{Degroote2010a, Moravveji2015b, Moravveji2016b, Szewczuk2018a, Buysschaert2018c, Szewczuk2021a, Pedersen2021a, Michielsen2021a, Bowman2021c*}, which span the mass range of $3 < M < 9$~M$_{\odot}$. These high-radial order g~modes are in the asymptotic regime \citep{UNNO_BOOK_2}, and can be identified using the morphology of their period spacing patterns, which are the period differences of consecutive radial order modes of the same degree expressed as a function of the mode period \citep{Miglio2008b, Bouabid2013}.

At higher masses between approximately $6 < M < 25$~M$_{\odot}$ are the \bcep pulsators, which pulsate in low-radial order pressure (p) and g modes with periods of order hours \citep{Lesh1978a, Sterken1993a, Stankov2005, Burssens2019a, Mozdzierski2019} driven by the $\kappa$ mechanism operating in the Fe-bump at 200\,000~K \citep{Dziembowski1993e, Pamyat1999b}. Despite the variability in early-B stars being known for over a century \citep{Frost1902}, their relative scarcity and the focus of planet-hunting space photometry missions towards low-mass stars, the availability of time series photometry of massive stars has been limited. Thanks to the all-sky TESS mission \citep{Ricker2015}, this obstacle to massive star asteroseismology has recently been removed (see e.g. \citealt{Pedersen2019a, Handler2019a, Bowman2019b, Bowman2020b, Burssens2020a}). However, the low-radial order pulsation modes of \bcep stars mean that the method of mode geometry identification by means of period spacing patterns is not typically possible, because their low-radial order pulsation modes are not usually in the asymptotic regime. Furthermore, time-series photometry is typically limited to detecting only low-degree (i.e. $\ell \leq 4$) modes because of geometric cancellation effects \citep{Dziembowski1977c, Daszy2002b, ASTERO_BOOK}. This represents a bias since typically only low-degree modes are then considered in subsequent forward asteroseismic modelling. Therefore, this means that another method of mode identification is usually needed to identify and fully explain the pulsation modes of \bcep stars.

Historically there has been great success in performing mode identification of slowly-rotating (i.e. $v\,\sin\,i \lesssim 50$~km\,s$^{-1}$) \bcep stars using rotationally-split modes (see e.g. \citealt{Handler2004b, Aerts2004b, DeRidder2004d}). But the ground-based data sets, both photometric and spectroscopic, of such systems typically constitute hundreds of hours of observations and require multi-site observations to produce a sufficient duty cycle suitable for asteroseismic modelling. Hence, there have only been a handful of \bcep stars as subjects of in-depth asteroseismic modelling, for example $\nu$~Eri \citep{Ausseloos2004, Daszy2010a}, 12~Lac \citep{Desmet2009b, Daszy2013b}, V836~Cen \citep{Aerts2003d, Dupret2004b}, and $\theta$~Oph \citep{Briquet2005, Daszy2009b}. These pioneering studies proved that the best constraints on massive star interiors come from a combined asteroseismic analysis of light curves and time-series spectroscopy. Specifically photometric and spectroscopic time series with a high cadence of order tens of minutes that adequately sample the pulsation periods and long durations of order months are needed. On the other hand, even long-term ground-based time series may not be sufficient because of gaps caused by the day-night cycle and poor weather. Alternatively, space-based telescopes providing continuous observations unhindered by the effects of the Earth's atmosphere are an ideal environment to assemble the necessary data to perform asteroseismology of massive stars. The ongoing TESS mission is providing the time-series photometry and the pulsation frequencies of many \bcep stars (e.g. \citealt{Burssens2020a}), but the time-series spectroscopy necessary for mode geometry identification in these stars is generally lacking.

In this paper we present the primary science case of massive star asteroseismology for the new Belgian-led European Space Agency (ESA) cubesat mission CubeSpec. In section~\ref{section: design}, we summarise the mission specifications and requirements for the primary science goal of the CubeSpec mission. In section~\ref{section: strategy}, we discuss the results of a literature study for pulsation and combine it with the analysis of recent time series photometry from the TESS mission to construct a prioritised target list for CubeSpec. In section~\ref{section: analysis}, we provide an example of the expectations of CubeSpec data for the future empowerment of massive star asteroseismology, and we conclude in section~\ref{section: conclusions}.


\section{CubeSpec mission science requirements}
\label{section: design}

Driven by the search for exoplanets, there has been great progress in assembling high-precision light curves of hundreds of thousands of stars thanks to the CoRoT \citep{Auvergne2009}, Kepler/K2 \citep{Borucki2010, Koch2010, Howell2014}, and TESS \citep{Ricker2015} missions. Clear advantages include not being hampered by the Earth's atmosphere, nor the daily interruptions caused by the day-night cycle. However, such missions are based on relatively large and expensive platforms, which require long development, testing and delivery phases, and consequently increased hardware and launch costs. Today, long-term light curves of bright stars can be assembled relatively easily using economical cubesats. A highly successful example of this is the BRITE-Constellation mission \citep{Weiss2014, Weiss2021a}. However, there remains an unfilled niche for developing a cost-effective cubesat designed to assemble time series spectroscopy. The CubeSpec mission is specifically designed to serve as a valuable proof-of-concept and in-orbit demonstration that low-cost and high-quality spectroscopy from space is feasible and cost effective.

The original design specifications of the CubeSpec cubesat were outlined by \citet{Raskin2018}. Starting from this initial mission design, a full preliminary design study was performed in the framework of the ESA General Support Technology Programme (GSTP) in-orbit demonstration cubesat programme, resulting in a successful preliminary design review in 2021. The mission has since started the implementation phase with launch planned in early 2024. The payload includes a Cassegrain telescope with a rectangular primary mirror with a collecting area of $9 \times 19$~cm$^{2}$, and a compact high-resolution echelle spectrograph with a resolving power of R~$\simeq 55\,000$. One of the engineering goals of CubeSpec is to demonstrate that the spectrograph design can be configured with minimal changes in hardware (e.g. detector, diffraction grating or prism disperser) or adjustment (e.g. change the angle of the diffraction grating to select a new wavelength range) to obtain $50 \lesssim {\rm R} \lesssim 50\,000$ over a wavelength range of $2000-10\,000~\AA$. Therefore, CubeSpec serves as a proof-of-concept of how high-quality spectroscopy can be delivered from a standard cubesat, with a typical timeline of going from science case to first in-orbit observations of approximately 1~yr. The design of the CubeSpec spectrograph offers significant flexibility and can be tuned to the needs of a specific science case. We focus here on the requirements of the primary science case, asteroseismology of massive stars, which was selected for the in-orbit demonstration mission.

For CubeSpec's primary science goal of delivering time series spectroscopy to enable massive star asteroseismology, there are several data requirements including the signal-to-noise ratio (SNR), spectral resolving power, cadence and monitoring time span. These science requirements define the technical specifications for the end-user to be able to measure LPV in the extracted 1D spectrum. Motivated by highly successful studies of massive star asteroseismology using ground-based time series spectroscopy (see e.g. \citealt{Aerts1992b, Aerts1998c, Aerts2004b, DeRidder2004d, Briquet2012}), the requirements of CubeSpec include are as follows: (i) a high-spectral resolving power of $R \geq 50\,000$ to detect and sufficiently sample rotationally broadened line profiles and characterise their temporal behaviour; (ii) a high continuum SNR~$>200$ in an individual spectrum to clearly identify LPV, which are of order a few per~cent of the light continuum; (iii) a high observational cadence allowing sufficient sampling of the pulsation frequencies, which are typically between $2 < P_{\rm puls} < 8$~hr in \bcep stars, such that at least 5-10 observations per pulsation period are obtained; (iv) a short integration time, maximally a few per~cent of the pulsation period, to avoid temporal smearing of the LPV; and (v) a long-duration time series to provide a precision on the pulsation frequency of order 1\,\%, such that a period of 1~d would require a time base of order 100~d.

The key spectroscopic diagnostic lines utilised for detecting pulsations in slowly-rotating \bcep stars are primarily the silicon triplet (Si\,III) at $\lambda\lambda$~4552, 4567, and 4574~$\AA$ \citep{ASTERO_BOOK}. Also, He~{\sc I} lines can be useful, especially for spectra with a low SNR ratio ($<100$) if the silicon triplet is too noisy. An extended wavelength coverage that includes Balmer, He~I, He~II lines and metal lines is also useful in the determination of effective temperatures and surface gravities of \bcep stars. Furthermore, a wavelength range that includes static lines produced by the interstellar medium provide an anchor-point for the wavelength calibration of CubeSpec spectra in addition to the on-board wavelength calibration source \citep{Raskin2018, Vandenbussche2021a}. Given these specifications and the technical design of the instrument, the wavelength range of CubeSpec is maximised to include from H$\gamma$ at $4340~\AA$ to the He~I line at $5876~\AA$ whilst meeting the required high spectral resolving power \citep{Vandenbussche2021a}.


\section{CubeSpec target list prioritisation}
\label{section: strategy}

The observational strategy for the primary science case of CubeSpec is to monitor a sample of a few \bcep stars for a minimum duration of 1 to 3 months\footnote{This is the minimum operational lifetime requirement of the CubeSpec mission, but the goal is to remain operational for at least 1~yr.}. The short orbital period (i.e. minimally 90~min) and anti-Solar viewing angle of CubeSpec's planned near-Earth orbit means that targets will only be viewable for approximately 50\,\% of each orbital period \citep{Vandenbussche2021a}. Two visits per 45-min window to the same target is possible, as is one visit for every $N$ orbits. This means to maximise the science potential of CubeSpec, an optimum target list and observing schedule that balances the necessary integration time and desired cadence for each high-priority \bcep star is needed prior to mission launch. 

\begin{figure}
\centering
\includegraphics[width=0.99\columnwidth]{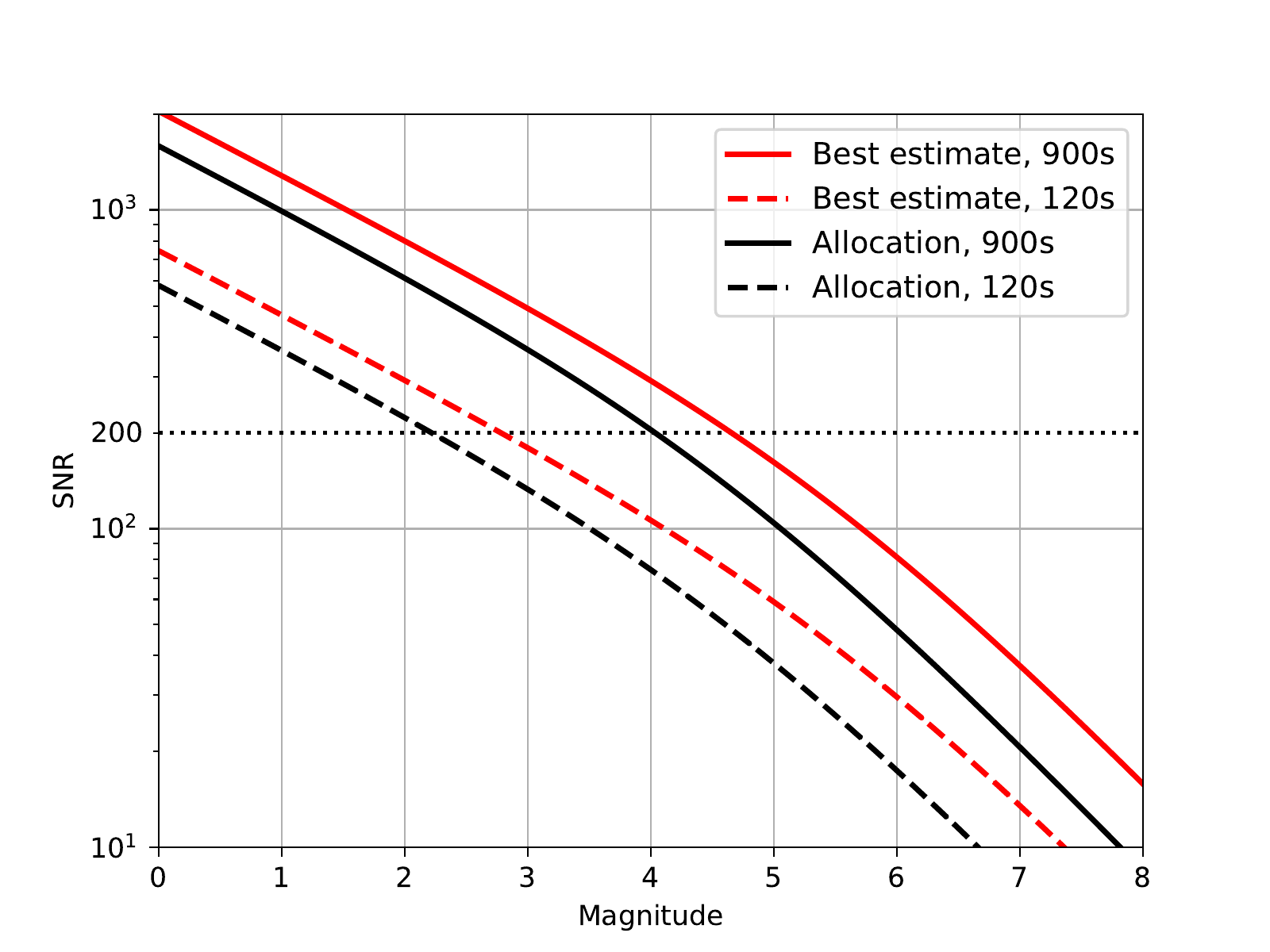}
\caption{CubeSpec signal-to-noise ratio (SNR) curves for a B0\,V star of different brightness in units of $V$~mag based on a conservative instrument performance allocation (black) and a best case performance scenario estimate (red) for exposure times of 120 and 900~sec. The SNR is estimated at a wavelength of $5000~\AA$ assuming $R = 55\,000$.}
\label{figure: exposure times}
\end{figure}

It is a mission requirement that at least three high-priority \bcep stars are monitored continuously during the minimum expected lifetime of the CubeSpec mission. The science requirement that each spectrum of each visit must have SNR > 200, coupled with the 45-min observing window limits the possible targets to bright $V \leq 4$~mag stars. This is because a $V = 4$~mag star of spectral type B0~V is expected to need an integration time of 15~min to reach SNR > 200 at the silicon triplet with CubeSpec's spectrograph \citep{Raskin2018}. Updated SNR curves based on the CubeSpec design are shown in Fig.~\ref{figure: exposure times}.

Ideally a larger sample of targets is required to serve as a pool to choose from in case the observing strategy needs to change post-launch for unforeseen technical reasons and to fill any additional observing time. For example, it is possible that a restriction in the attitude range towards more benign solar aspect angles may be needed for thermal reasons. Moreover, some targets will only be viewable for a fraction of each orbital period based on their sky location. As part of the CubeSpec design phase, several secondary science cases were identified that would benefit from obtaining high-resolution spectroscopic time series observations using either the current high-resolution design of CubeSpec or one with a lower resolving power and larger wavelength coverage \citep{Raskin2018}. These secondary science cases include: asteroseismology of solar-type stars; absolute flux calibration of stellar atmosphere models; diffuse interstellar bands; and exoplanet and stellar host activity. We do not expand upon these further in the current paper, but available observing time after observing high-priority \bcep targets will be allocated to the secondary science goals of CubeSpec \citep{Raskin2018}.


	\subsection{SIMBAD search for early-B stars}
	\label{subsection: simbad}
	
\begin{figure}
\centering
\includegraphics[width=0.99\columnwidth]{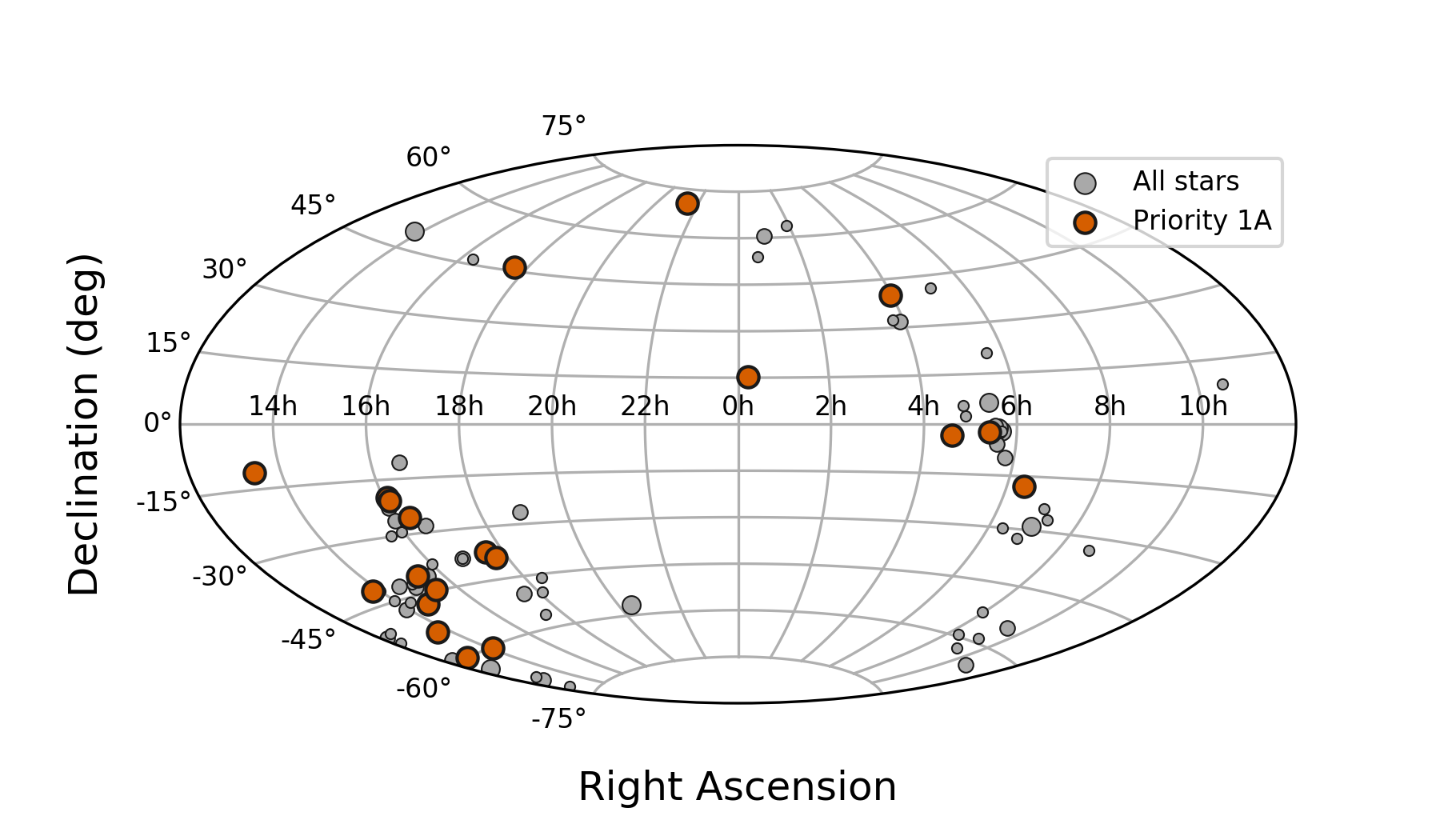}
\includegraphics[width=0.99\columnwidth]{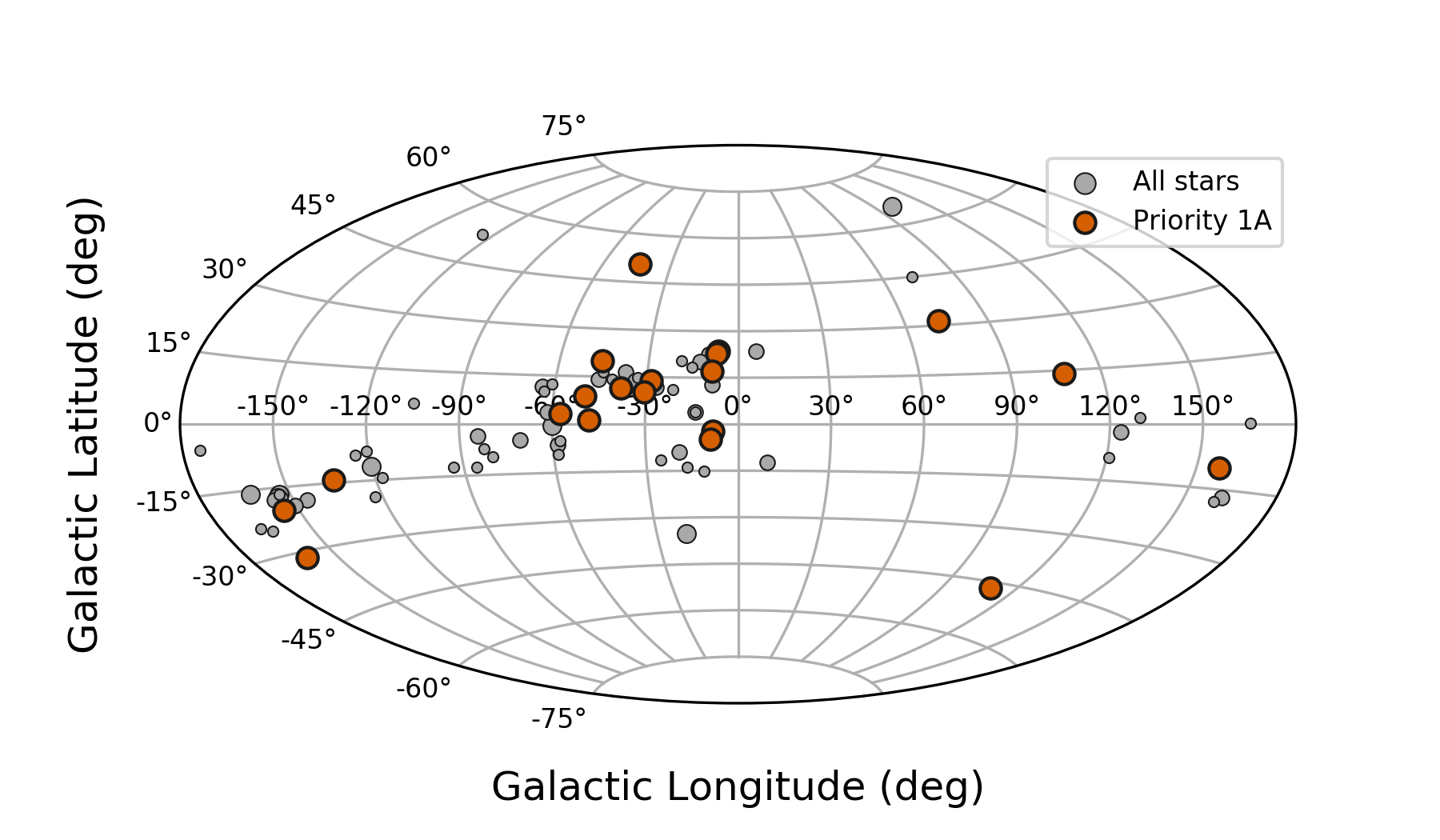}
\caption{Sky distribution of potential CubeSpec mission targets, as listed in Table~\ref{table: targets}.}
\label{figure: sky}
\end{figure}
	
	To find suitable \bcep stars for CubeSpec, we combine a literature study of previous detections of pulsation (e.g. \citealt{Sterken1993a, Aerts2003a}), and new TESS photometry of bright stars. We use {\sc simbad} to search for bright ($V$~$\leq 4$~mag) stars with spectral types between O9 and B3 for all luminosity classes, because this is the typical spectral type range for \bcep stars \citep{Lesh1978a, Sterken1993a, Stankov2005}. All the stars returned from this search are given in Table~\ref{table: targets}, in which we include the name of the star, the TESS input catalogue (TIC) identification number, coordinates, $B$ and $V$ magnitudes and spectral type from {\sc simbad}. For each target that has previously been identified as a \bcep pulsator and its pulsations analysed in detail, we include a reference in Table~\ref{table: targets} giving priority to stars that have been the subject of previous LPV and photometric studies. For targets that have not been the subject of a detailed LPV study, we include the projected rotational velocity from the fifth version of the Bright Star Catalogue (BCS5; \citealt{Hoffleit1991}), or from catalogue papers \citep{Abt2002, Rivinius2003b, Wolff2007}. We also cross-matched the list of targets with the SB9 binary catalogue \citep{Pourbaix2004}, and include a column in Table~\ref{table: targets} to indicate if a target is known to be a member of a binary or multiple system. 
	
	The distribution of the targets in Table~\ref{table: targets} in the sky is shown in Fig.~\ref{figure: sky}. The majority of the high priority (1A) \bcep stars identified from our search are generally located close to the galactic plane but are reasonably well spread in terms of right ascension (see sections \ref{subsection: priorities} and \ref{subsection: priority1} for definition of priority 1A). This allows for a more detailed mission profile study and realistic target visibilities to be calculated in future work. The dominant pulsation amplitudes confirmed by spectroscopic LPV studies among priority 1A stars ranges from 1 to 40~km\,s$^{-1}$, with projected rotational velocities being between $1 \leq v\,\sin\,i \leq 170$~km\,s$^{-1}$.


	\subsection{TESS light curve analysis}
	\label{subsection: TESS}
	
	We also utilise the high-precision photometry from the all-sky TESS mission \citep{Ricker2015} to ensure a more complete photometric search for \bcep stars amongst our targets. We cross-match the list of stars in Table~\ref{table: targets} with the TESS archive to obtain available photometry. In some cases, the same TIC number is returned for some unresolved close-binary systems listed in Table~\ref{table: targets}, for example $\beta$~Sco, $\zeta$~Ori, $\eta$~Ori and $\alpha$~Cru. In cases such as these, but generally for all bright stars observed by TESS, the extracted light curve includes signal from multiple stars including background contaminants owing to the 21~arcsec TESS pixel size. Therefore, care is needed in claiming the detection of \bcep pulsations based solely on a TESS light curve of a bright star. If, for example, a fainter pulsating A-type star (i.e. $\delta$~Sct; \citealt{Breger2000b, Bowman2018a}) star is within the mask, pulsation modes similar in frequency to those of a \bcep star may be detected at significant amplitude. A \dsct star contributing $\sim 1$\,\% of the total light within the aperture with pulsation mode amplitudes of order 10~mmag would be diluted to have amplitudes of order 0.1~mmag and mimic low-amplitude \bcep pulsations. 

	Cycle 4 of the TESS mission is currently underway as of July 2021, and almost all bright O- and B-type stars were prioritised for short-cadence (i.e. 2~min) observations thanks to two successful Guest Investigator proposals which cover both the northern and southern ecliptic hemispheres (GO3059 and GO4074; PI: Bowman\footnote{\url{https://heasarc.gsfc.nasa.gov/docs/tess/approved-programs.html}}). However, not all stars in Table~\ref{table: targets} have been observed by TESS, because the observing strategy of TESS prior to cycle 4 avoided the ecliptic. Moreover, some stars fall between the gaps of the CCDs and the cameras, and some stars have not been observed to date but will be for the first time in the coming year. 

	For each target from our preliminary search given in Table~\ref{table: targets}, we download the simple aperture photometry (SAP) and pre-data search conditioning (PDC-SAP) TESS light curves for all available sectors from the Mikulski Archive for Space Telescopes (MAST\footnote{\url{https://archive.stsci.edu/}}). These light curves are extracted using the {\sc SPOC} pipeline \citep{Jenkins2016b}, and are available for many of the stars listed in Table~\ref{table: targets}, but not all (yet) because the the TESS mission is still ongoing. We check the assigned aperture masks are reasonable and convert the light curves into units of stellar magnitudes and perform additional detrending in the form of a low-order polynomial for each sector. We also extract our own light curves using simple aperture photometry routines based on a threshold criterion for mask selection, which utilise the {\sc TESScut} \citep{Brasseur2019a} and {\sc Lightkurve} software packages \citep{Lightkurve2018}. Our light curves include background correction based on selecting pixels below the chosen threshold and perform a principal component analysis (PCA) to remove systematics (see Garcia et al., submitted to A\&A). Using also our own approach, we can make a comparison to the available 2-min PDC-SAP light curves.

	The majority of stars in our sample are quite bright for TESS and are moderately or even heavily saturated. With brighter and more saturated stars, the number of pixels included as part of a simple aperture photometry light curve extraction method increases greatly. As a consequence this increases the likelihood of including nearby pulsating stars as contaminating sources in the light curve. In a heavily saturated scenario, simple aperture photometry is not a viable light curve extraction method if the saturated pixel columns extend to the edge of the CCDs. Hence alternative methods such as halo and smear photometry are preferable, and have been successfully applied to {\it Kepler} and K2 mission photometry (see e.g. \citealt{Pope2016a, Pope2019b, Pope2019c, White2017b}). Therefore, we make use of sectors in which the downloaded target pixel files include sufficient pixels, including saturated ones, to apply simple aperture photometry. Given that the goal of this work is to identify high-amplitude \bcep stars for the CubeSpec mission, we leave the detailed analysis of all individual TESS sectors to future work and mark the problematic stars in Table~\ref{table: targets}. 
	
	To aid in classifying the variability of each target, we calculate amplitude spectra using a discrete Fourier transform (DFT; \citealt{Kurtz1985b}) up to the Nyquist frequency of a light curve, which is 360~d$^{-1}$ for the 2-min short-cadence light curves, 24~d$^{-1}$ for the 30-min extracted light curves from TESS full frame images (FFIs) in cycles 1 and 2, and 72~d$^{-1}$ for the 10-min extracted light curves from FFIs in cycles 3 and 4. The typical frequency range of pulsation modes in \bcep stars are well sampled even with the longest available cadence of 30~min, but we prioritise short-cadence light curves to reduce the impact of amplitude suppression of pulsation modes because of longer integration times (see e.g. \citealt{Bowman_BOOK}). Variability in massive star light curves is diverse and common, with earlier studies of TESS light curves of O and B stars demonstrating that binarity, rotation, winds and pulsations are commonly observed \citep{Bowman2019b, Pedersen2019a, Burssens2020a}.


	\subsection{Prioritisation of targets}
	\label{subsection: priorities}

	For each star, the available literature and the amplitude spectrum of the TESS light curve is used to visually classify the dominant variability and assign a priority for CubeSpec. Since the uncertainties in stellar models propagate with age and the most successful asteroseismic modelling of \bcep stars were main sequence single stars (see review by \citealt{Bowman2020c}), we assign the highest priority to high-amplitude dwarf and giant \bcep stars. The priority classes used to classify all the stars in Table~\ref{table: targets} are: (1) \bcep stars; (2) stars with high-amplitude g~modes; (3) stars dominated by stochastic low-frequency variability (SLF) variability; and (4) other dominant variability (e.g. binarity, rotation). Furthermore, we sub-divide priority~1 stars into three categories: priority~1A are well-established \bcep stars based on LPV and time series photometry studies; priority~1B stars are those with confirmed LPV but the periods and amplitudes have not been measured, usually due to insufficient data coverage; and priority~1C are stars that show evidence of \bcep pulsations in their TESS light curves extracted as part of this work, but spectroscopic confirmation of LPV is lacking in the literature. 
	
	Also in Table~\ref{table: targets}, we include stars identified as non-radial g-mode pulsators, such as SPB and pulsating classical Be stars \citep{Waelkens1991c, Rivinius2013c}, as priority~2. Such stars are useful to study using time series spectroscopy for the purpose of asteroseismology (see e.g. \citealt{Rivinius2003b}). However, SPB stars and pulsating Be stars are typically rapid rotators, which complicates the analysis of LPV in time series spectroscopy. The other types of variability included in priority classes 3 and 4 are interesting to study using time-series spectroscopy, for example, SLF variability \citep{Bowman2019a, Bowman2019b, Bowman2020b}, wind variability, binarity and rotational modulation (e.g. \citealt{Rivinius1997b, Prinja2004a, Lefevre2005a, Simon-Diaz2017a, Simon-Diaz2018a}). However these different variability phenomena are not the primary science goal of the CubeSpec mission. Instead, CubeSpec is a proof-of-concept mission to demonstrate that high-resolution time series spectroscopy can be achieved from space for a relatively low cost. This means to adequately benchmark the data, we prioritise \bcep stars that have studied before as opposed to searching for variability in previously unknown pulsators.
	
	Table~\ref{table: targets} further represents the results of an up-to-date literature study of pulsation in bright late-O and early-B stars. This magnitude-limited search for variability is enhanced and updated thanks to our analysis of new TESS light curves.


	\subsection{Priority~1 stars: \bcep pulsators}
	\label{subsection: priority1}

	For stars that are assigned priority~1A based on our literature search and have available TESS photometry, we provide the light curves and amplitude spectra in Figs~\ref{figure: priority1_1} and \ref{figure: priority1_2}. In these figures, amplitude spectra shown in black correspond to stars for which the PDC-SAP light curve is deemed superior (i.e. fewer systematics and higher SNR of pulsations) compared to our own extracted light curves from the FFIs, and those in blue are for which the quality of our extracted light curve from the FFIs is deemed superior. Of course, not all stars included within priority~1A can be observed continuously as part of the CubeSpec mission, so we prioritise stars based on their brightness, low rotational velocity and high pulsation amplitudes. Brighter stars require shorter integration times, and lower rotational velocities and higher pulsation amplitudes allow LPV to be more easily characterised since such stars have spectral lines that are dominated by pulsational broadening as opposed to rotational broadening. Although there is not a one-to-one relationship between the amplitude of pulsation modes in photometry and spectroscopy, it is generally the case that the pulsation amplitudes correlate (see e.g. \citealt{Aerts2003a}). 
	
	Another important factor to consider is the range of pulsation frequencies, since higher frequencies (i.e. shorter periods) require a higher cadence for sufficient sampling following the Nyquist theorem. As an example, stars with pulsation mode frequencies above approximately 8~d$^{-1}$ would require two visits for each 45-min window for each 90-min CubeSpec orbit, in order to produce a higher effective Nyquist frequency compared to that from only one visit per orbit. Therefore, \bcep stars with pulsation mode frequencies below 7~d$^{-1}$ are preferable as only a single visit per CubeSpec orbit would be needed. In Table~\ref{table: targets}, we provide the pulsation mode frequency range of priority~1A stars, which is based on the literature and our preliminary analysis of the new TESS data. None of the stars assigned priority~1A are newly discovered \bcep stars; all were known to pulsate prior to the TESS mission. In the case of stars assigned priority 1B and 1C, it varies from star to star on whether they were previously known to be \bcep stars. Priority 1B stars have been the subject of successful LPV studies, whereas priority 1C stars have not been and can be considered newly-discovered candidate \bcep stars based on our analysis of the available TESS data. 
			
	All the stars assigned priority~1A in Table~\ref{table: targets} would be excellent targets for the CubeSpec mission, with long-term and continuous optical time-series spectroscopy undoubtedly revealing new insight of their atmospheres and interiors. The fact that some targets have been studied before does not exclude them from the CubeSpec mission, since \bcep stars are known to exhibit significant amplitude and frequency modulation in their pulsation modes over long time spans \citep{Degroote2009a, Degroote2013}. Furthermore, some \bcep stars have also been shown to exhibit period changes that are consistent with stellar evolution \citep{Jerzykiewicz1999a, Neilson2015b}. Therefore, \bcep stars that have been previously studied are valuable to re-visit with the CubeSpec mission for a variety of astrophysical reasons in addition to the main goal of performing forward asteroseismic modelling. Moreover, the data gathered in the literature to study LPV in \bcep stars typically consists of a few hundred spectra spanning a few days or a few weeks. In rare cases are thousands of spectra spanning months gathered for such a star, as such a data set is extremely expensive and not generally considered a competitive proposal for time allocation committees for ground-based telescopes.
	
	To narrow down the options from priority~1A to the necessary 3-8 stars with a total estimated integration time of 45~min per orbit, we select the \bcep stars with the highest asteroseismic potential based on the known frequency range and amplitudes of their pulsations and each star's projected rotational velocity. \bcep stars with low projected rotational velocities and large pulsation amplitudes have line profiles that are typically dominated by pulsational broadening as opposed to rotational broadening (see e.g. \citealt{Aerts1992b}), and make excellent targets for CubeSpec. These high-potential targets for asteroseismology based on CubeSpec data are provided in Table~\ref{table: final targets} and are highlighted in the sky distribution plot in Fig.~\ref{figure: sky}. For each star we also provide best case and conservative estimates of the necessary CubeSpec exposure time required to achieve SNR~$>200$ at $5000~\AA$. The reason for the difference between the conservative allocation and best-case exposure times is that the SNR has been estimated based on the blaze peak of the spectrograph. At the edge of the free-spectral range, efficiency drops to approximately 40\,\% of the blaze peak for the bluer echelle orders, but it is anticipated that this lost flux can be recovered in the previous or next order \citep{Raskin2018, Vandenbussche2021a}. Hence the allocation exposure times are a ``worst case'' scenario in this respect. 
	
	In our selection of the best \bcep stars as targets for CubeSpec, we opt to prioritise slowly rotating stars with large pulsation amplitudes, as these stars are predicted to have spectral line broadening that is dominated by pulsations instead of rotation. The analysis of such stars typically yields the identification of low-angular degree modes \citep{Aerts1992b, Briquet2003a}. Since the ultimate scientific goal of the CubeSpec mission is to provide data to perform forward asteroseismic modelling, we prioritise slowly-rotating \bcep stars because the low-angular degree modes in such stars have often been more successfully utilised as inputs to forward asteroseismic modelling. Moreover, calculating stellar structure and evolution models, using for example {\sc MESA} \citep{Paxton2011, Paxton2019}, and theoretical pulsation mode frequencies, using for example {\sc GYRE} \citep{Townsend2013b}, for the purpose of forward asteroseismic modelling is more robust for slowly rotating stars.
	
	In total, the eight targets listed in Table~\ref{table: final targets} amount to a total of exposure time of 19 mins in the best case scenario, such that each target could be observed twice per 45-min window, and a total of 35 mins in the conservative allocation performance scenario. We note that this estimate does not include CCD read out time, which is estimated to be of order seconds for a single exposure. Nor does it include pointing time and relaxation time, which is anticipated to be a few minutes per pointing. Hence there is ample time to observe 8 \bcep stars at SNR~$\geq 200$ per CubeSpec orbit, and have remaining time for secondary science goals or additional exposures of some of the \bcep stars. Future work on mission profile analysis based on the CubeSpec's orbit and target visibilities will reveal the specific windows of visibility for each of the priority targets.

\begin{table}
\caption{Targets of highest priority for the primary science case of the CubeSpec mission.} 
\begin{center}

\resizebox{0.99\columnwidth}{!}{

\begin{tabular}{c c c c c c c}
\hline
\multicolumn{1}{c}{Name} & \multicolumn{1}{c}{$V$} & \multicolumn{1}{c}{Sp. Type} & \multicolumn{1}{c}{Puls. Freq} & \multicolumn{1}{c}{$A_{\rm rad\,vel}$} & \multicolumn{1}{c}{Exp. Time} \\
\multicolumn{1}{c}{} & \multicolumn{1}{c}{(mag)} & \multicolumn{1}{c}{} & \multicolumn{1}{c}{(d$^{-1})$} & \multicolumn{1}{c}{(km\,s$^{-1}$)} & \multicolumn{1}{c}{(s)} \\
\hline
$\gamma$~Peg	&	2.84	&	B2\,IV  			&	6.59		&	3.5		&	130 / 230	\\        
$\nu$~Eri			&	3.93	&	B2\,III          		&	[5.62, 5.76]&	$<20$	&	400 / 790	\\
$\beta$~CMa		&	1.97	&	B1\,II-III       		&	[3.97, 4.18]&	$<11$	&	60 / 100	\\
$\beta$~Cru		&	1.25	&	B1\,IV           		&	[5.23, 5.96]&	$<3$		&	30 / 50	\\
$\alpha$~Lup		&	2.29	&	B1.5\,III       		&	[3.85, 4.22]&	$<7$		&	80 / 130	\\
$\beta$~Sco		&	2.50	&	B1\,V + B2\,V		&	5.77		&	$<14$	&	100 / 200	\\
$\sigma$~Sco		&	2.89	&	O9.5\,V + B7\,V		&	[4.05, 4.17]&	$<40$	&	130 / 230	\\ 
$\beta$~Cep		&	3.23	&	B0.5\,III       		&	[4.92, 5.42]&	$<21$	&	190 / 350	\\
\hline
\end{tabular} }
\tablefoot{Columns are star name, $V$~mag, spectral type from {\sc simbad}, the indicative pulsation frequency and radial velocity amplitude detected in LPV studies, and the estimated CubeSpec exposure time needed to achieve SNR~$>200$ at $5000~\AA$ assuming best/conservative operating conditions. Further details for each target are provided in Table~\ref{table: targets}.}
\end{center}
\label{table: final targets}
\end{table}


\section{Analysis method of future CubeSpec data}
\label{section: analysis}

The pulsations generated inside a star cause geometrical and time-dependent deformations in the line forming region near the stellar surface and cause LPV of a pulsating star's spectral lines. The geometry of a pulsation mode defines the morphology of the variability near a star's surface and the frequency of the perturbation to a spectral line. To disentangle the geometries of the multi-periodic pulsations in massive stars, a high-quality spectroscopic time series with a spectral resolving power, $R = \lambda/\delta\lambda \geq 50\,000$ and SNR~$\geq 200$ is required to perform robust mode geometry identification \citep{Aerts1992b, Briquet2003a}. So far this has only been achieved using ground-based telescopes and these short-term observations suffer from large daily time gaps in the case of single-site observations, which limit the ability to disentangle different pulsation frequencies, hence limit the scientific value of the data. 

For \bcep stars, there are two main types of pulsation modes to consider: p and g modes. The former have predominantly radial perturbations to the line profile \citep{Aerts2003a}, whereas g modes produce predominantly tangential velocity perturbations to the line profile \citep{DeCat2002c}. There are other physical mechanisms that cause LPV in early-type stars. One example is stellar winds \citep{Puls2008c, Sundqvist2011b}, but these are typically a negligible contribution to LPV compared to the effect of pulsations in early B-type dwarf stars, which have relatively low mass-loss rates \citep{Bjorklund2021a}. Moreover, LPV caused by stellar winds is more stochastic and can be decoupled from the periodic effect of coherent pulsation modes.

As a demonstration of what we expect from CubeSpec, we use the {\sc bruce}\footnote{\url{http://www.astro.wisc.edu/~townsend/static.php?ref=bruce}} code \citep{Townsend1997, Townsend_PhD} to simulate a time-series of line profile variations due to the effect of stellar pulsations. Although {\sc bruce} allows to account for the effects of stellar rotation (using the traditional approximation) and gravity darkening, these are ignored in the present simulations for simplicity. The output of the {\sc bruce} code is in the form of a time-series of perturbed due to (non-)radial stellar pulsations models, where each mesh point on the stellar surface contains information about local velocity, gravity, surface area, temperature, et cetera. For simplicity and because temperature and surface area perturbations due to (non-)radial pulsations are a second-order effect compared to the velocity variations \citep{Townsend1997, Townsend_PhD}, the two former effects are neglected in these simulations. We use the {\sc SynthV} line formation code \citep{Tsymbal1996} to compute a library of intrinsic spectral line profiles at different positions on the stellar disk and take into account all intrinsic broadening mechanisms. A local profile corresponding to the physical properties of each surface element computed by {\sc bruce} is then obtained by interpolating within that library of intrinsic line profiles. Ultimately, we integrate over the visible part of the surface to give a disk-integrated line profile at a particular time stamp. 

\begin{figure}
\centering
\includegraphics[width=0.99\columnwidth]{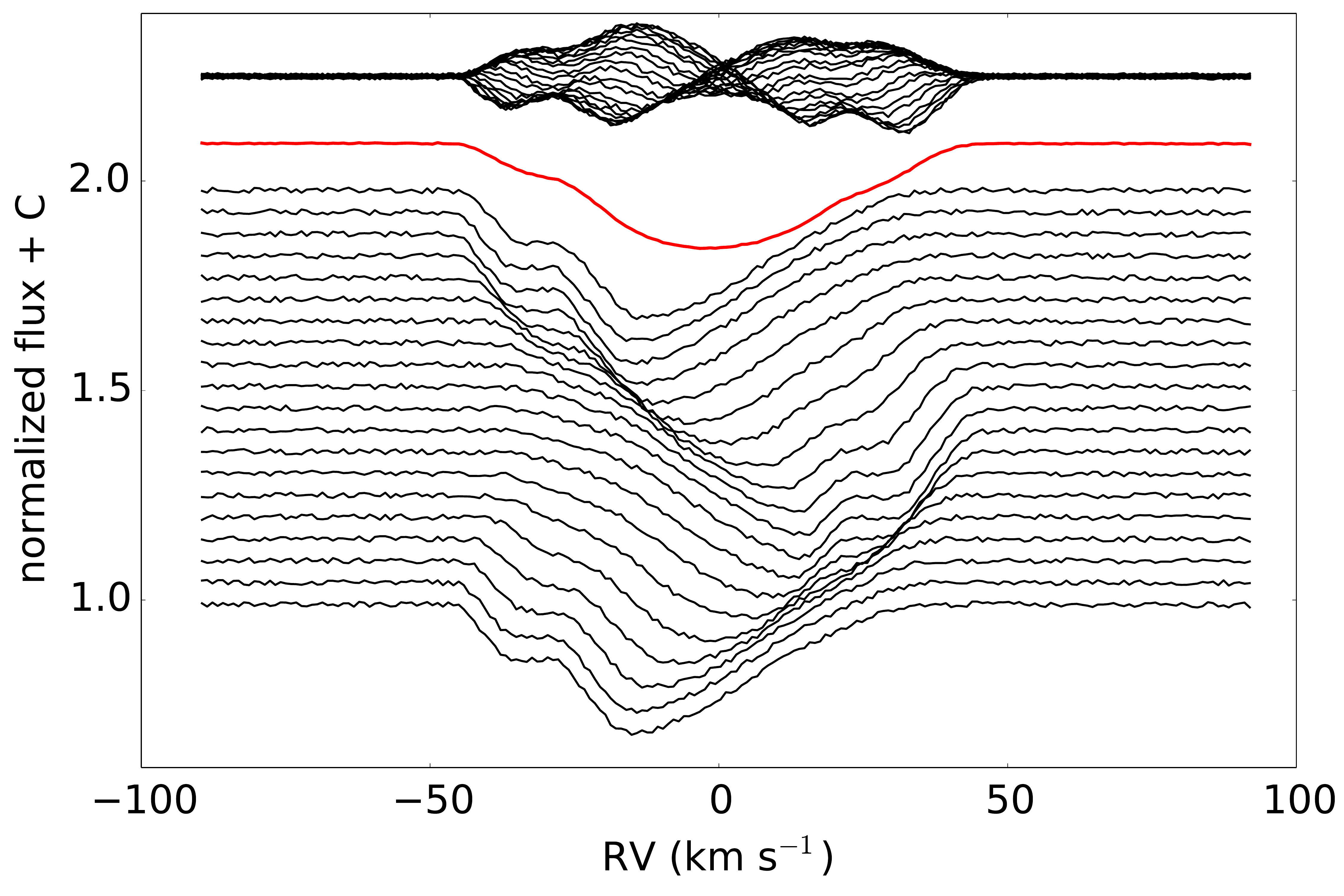}
\caption{Simulated LPV for one pulsation cycle of \bcep using the {\sc bruce} code assuming a resolving power of 55\,000, an inclination angle of 60$^{\circ}$, a projected rotational velocity $v\,\sin\,i$ of 25~km\,s$^{-1}$, with a cadence of 15~min and noise to emulate a signal-to-noise ratio of 200. The red line shows the average profile, and at the top the residuals obtained after subtracting the average profile are shown.}
\label{figure: bruce}
\end{figure}

As an example, we use $\beta$~Cep and its dominant radial pulsation mode of 5.2497740~d$^{-1}$ \citep{Aerts1994b} as a prototype to create a ``toy model’’ of line profile variations that the CubeSpec mission will provide given its expected performance. We use the pulsation amplitude of 21~km\,s$^{-1}$, inclination angle of the rotation axis of 60$^{\circ}$, and projected rotational velocity $v\,\sin\,i$ of 25~km\,s$^{-1}$ as deduced by \citet{Aerts1994b} from the high-resolution spectra obtained with the Aurelie spectrograph. We simulate one complete pulsation cycle at a cadence of 15~min as an example, and show the resulting LPV along with the residuals obtained after subtracting the average profile in Fig.~\ref{figure: bruce}, with each spectrum including noise to emulate SNR~$= 200$. The LPV caused by pulsations is clearly seen.

In the analysis of a spectroscopic time series of a pulsating star, there are generally two methods that can be used: (i) the moment method \citep{Balona1986a, Balona1986c, Balona1987a, Aerts1992b, Aerts1996b, Briquet2003a}, and (ii) the pixel-by-pixel method \citep{Schrijvers1997, Mantegazza2000b, Zima2006b, Zima2006c}. The moment method requires the numerical integration of the statistical moments of an observed spectral line and describes LPV in terms of: equivalent width (0th moment); centroid velocity (1st moment); profile width (2nd moment); and profile skewness (3rd moment). Such a method is most effective for slowly-rotating stars (i.e. $v\,\sin\,i \lesssim 50$~km\,s$^{-1}$), because the rotation period is therefore much larger than the pulsation period. Thus, the dominant line broadening mechanism is because of pulsations as opposed to rotation. The analysis of slowly rotating \bcep stars typically leads to the identification of low-angular degree pulsation modes, which are ideal observational inputs for forward asteroseismic modelling. On the other hand, the pixel-by-pixel method is sensitive to the phase and amplitude of a pulsation mode as a function of the line profile. It is more suitable for moderately and rapidly rotating stars (i.e. $v\,\sin\,i \gtrsim 50$~km\,s$^{-1}$), and typically leads to the identification of high-angular degree pulsation modes, which are seen as bumps moving through the spectral line profile. However, the pixel-by-pixel method has the disadvantage that inference from it is limited by the signal-to-noise of the spectrum. Therefore, since we have optimised the CubeSpec target list for slowly rotating stars with pulsationally dominated spectral line broadening to identify low-angular degree pulsation modes, we will primarily apply the moment method to maximise the feasibility of future forward asteroseismic modelling.

Both the moment and pixel-by-pixel methods are available within the {\sc FAMIAS} software package \citep{Zima2008a, Zima2008c}, which is commonly used for the purpose of mode geometry identification from spectroscopic time series. We do not analyse the simulated LPV generated using the {\sc bruce} code, as this is beyond the scope of the current paper and because it is only a toy model. For recent applications of the {\sc FAMIAS} software package for identifying the mode geometry of pulsating stars, see \citet{Brunsden2018a}, \citet{Aerts2019a}, \citet{Mozdzierski2019} and \citet{Johnston2021a}.

CubeSpec mission data will be made public to the wider community by the end of the nominal mission. In this way, a legacy archive hosted at the Institute of Astronomy, KU Leuven will be created to host these data for follow-up and complementary science cases.


\section{Summary}
\label{section: conclusions}

In this paper, we have summarised the technical specifications and primary science requirements of the upcoming Belgian-led ESA CubeSpec in-orbit demonstration and space mission, which will assemble high-resolution optical spectroscopic time series of bright massive stars for the purpose of pulsation mode geometry identification and asteroseismology. We have performed an extensive literature search for \bcep pulsators which span the spectral type range of O9 to B3 and are brighter than $V \leq 4$~mag. We also analyse new TESS mission photometry and combine this with our literature study to prioritise targets with the highest asteroseismic potential for future forward asteroseismic modelling. Considering the science requirements and technical specifications of the CubeSpec mission, we identify $\gamma$~Peg, $\nu$~Eri, $\beta$~CMa, $\beta$~Cru, $\alpha$~Lup, $\beta$~Sco, $\sigma$~Sco and $\beta$~Cep as high-priority targets (see Table~\ref{table: final targets}). Our full target list is provided in Table~\ref{table: targets}, and serves as a useful catalogue of pulsation amongst bright early-B stars.

Based on conservative estimates of exposure, read-out and slew time, we expect sufficient time for at least one visit for each of these eight \bcep stars per orbital period assuming a minimum orbital period of 90~min for CubeSpec. In a best-case scenario, in which the instrumental design performs well, we anticipate ample time to observe each of these eight \bcep stars twice per orbital period and have time left for secondary science goals.  We have also generated a simplistic toy model based on the known properties of \bcep to demonstrate the feasibility of detecting LPV using CubeSpec's expected performance and mission specifications. The LPV detected in the extracted 1D spectrum of massive stars will be analysed using the KU Leuven in-house code {\sc FAMIAS} \citep{Zima2008a, Zima2008c}. CubeSpec is currently on schedule to be launched and operational in-orbit in 2024, and provide excellent data for massive star asteroseismology soon thereafter.


\begin{acknowledgements}

The CubeSpec mission development is funded by the Belgian federal Science Policy Office (BELSPO) through the in-orbit demonstration ESA General Support Technology Programme (GSTP) programme of the European Space Agency.

The TESS data presented in this paper were obtained from the Mikulski Archive for Space Telescopes (MAST) at the Space Telescope Science Institute (STScI), which is operated by the Association of Universities for Research in Astronomy, Inc., under NASA contract NAS5-26555. Support to MAST for these data is provided by the NASA Office of Space Science via grant NAG5-7584 and by other grants and contracts. Funding for the TESS mission is provided by the NASA Explorer Program. 

The authors are grateful to the referee for the positive report.

DMB gratefully acknowledges a senior postdoctoral fellowship from the Research Foundation Flanders (FWO) with grant agreement no.~1286521N.

The research leading to these results has (partially) received funding from the KU~Leuven Research Council (grant C16/18/005: PARADISE), from the Research Foundation Flanders (FWO) under grant agreement G0H5416N (ERC Runner Up Project), and from the BELgian federal Science Policy Office (BELSPO) through PRODEX grant PLATO.

TVR gratefully acknowledges support from the Research Foundation Flanders (FWO) under grant agreement number 12ZB620N.

This research has made use of the {\sc SIMBAD} database, operated at CDS, Strasbourg, France; the SAO/NASA Astrophysics Data System; and the VizieR catalog access tool, CDS, Strasbourg, France.

This research has made use of the \texttt{PYTHON} library for publication quality graphics (\texttt{MATPLOTLIB}; \citealt{Matplotlib_2007}), \texttt{Seaborne} \citep{Seaborn_2021},  \texttt{Numpy} \citep{Numpy_2006, Numpy_2011, Numpy_2020}.

\end{acknowledgements}


\bibliographystyle{aa}
\bibliography{/Users/dominic/Documents/RESEARCH/Bibliography/master_bib}


\begin{appendix}

\section{Target list table}
\label{section: appendix: tables}

The extended target list of CubeSpec is given in Table~\ref{table: targets}, in which available variability classifications and priority categories based on available TESS data are provided.

\onecolumn
\tiny

\longtab{
\begin{landscape}

\begin{longtable}{c r r r c c c c c c c l c c}
\caption{\label{table: targets} Full target list of bright massive stars for the primary science case of the CubeSpec mission.} \\
\hline\hline
\multicolumn{1}{c}{Name} & \multicolumn{1}{c}{TIC} & \multicolumn{1}{c}{RA} & \multicolumn{1}{c}{dec} & \multicolumn{1}{c}{$B$} & \multicolumn{1}{c}{$V$} & \multicolumn{1}{c}{Sp. Type} & \multicolumn{1}{c}{$v\,\sin\,i$} & \multicolumn{1}{c}{bin} & \multicolumn{1}{c}{Puls. Freq.} & \multicolumn{1}{c}{$A_{\rm rad\,vel}$} & \multicolumn{1}{c}{LPV study reference} & \multicolumn{1}{c}{TESS var.} & \multicolumn{1}{c}{Notes}\\
\multicolumn{1}{c}{} & \multicolumn{1}{c}{} & \multicolumn{1}{c}{} & \multicolumn{1}{c}{} & \multicolumn{1}{c}{(mag)} & \multicolumn{1}{c}{(mag)} & \multicolumn{1}{c}{} & \multicolumn{1}{c}{(km\,s$^{-1}$)} & \multicolumn{1}{c}{} & \multicolumn{1}{c}{(d$^{-1}$)} & \multicolumn{1}{c}{(km\,s$^{-1}$)} & \multicolumn{1}{c}{} & \multicolumn{1}{c}{} \\
\hline
\endfirsthead
\caption{\it continued.}\\
\hline\hline
\multicolumn{1}{c}{Name} & \multicolumn{1}{c}{TIC} & \multicolumn{1}{c}{RA} & \multicolumn{1}{c}{dec} & \multicolumn{1}{c}{$B$} & \multicolumn{1}{c}{$V$} & \multicolumn{1}{c}{Sp. Type} & \multicolumn{1}{c}{$v\,\sin\,i$} & \multicolumn{1}{c}{bin} & \multicolumn{1}{c}{Puls. Freq.} & \multicolumn{1}{c}{$A_{\rm rad\,vel}$} & \multicolumn{1}{c}{LPV study reference} & \multicolumn{1}{c}{TESS var.} & \multicolumn{1}{c}{Notes}\\
\multicolumn{1}{c}{} & \multicolumn{1}{c}{} & \multicolumn{1}{c}{} & \multicolumn{1}{c}{} & \multicolumn{1}{c}{(mag)} & \multicolumn{1}{c}{(mag)} & \multicolumn{1}{c}{} & \multicolumn{1}{c}{(km\,s$^{-1}$)} & \multicolumn{1}{c}{} & \multicolumn{1}{c}{(d$^{-1}$)} & \multicolumn{1}{c}{(km\,s$^{-1}$)} & \multicolumn{1}{c}{} & \multicolumn{1}{c}{} \\
\hline
\endhead
\hline
\endfoot

\multicolumn{14}{l}{Priority~1A: confirmed \bcep stars with pulsation analysis based on LPV in the literature} \\
$\gamma$~Peg	&	51942308\tablefootmark{$\dagger$}		&	00:13:14.15&	15:11:00.94	&	2.61	&	2.84	&	B2\,IV  			&	5	&	$\checkmark$	&	6.59		&	3.5		&	Y \citep{Butkovskaya2007} 	&	$-$	&	{\,}\tablefootmark{1} \\              
$\epsilon$~Per	&	187107884	&	03:57:51.23&	40:00:36.78	&	2.71	&	2.89	&	B1.5\,III        		&	130		&	$\checkmark$	&	[5.30, 10.59]&	$<1$		&	Y \citep{DeCat2000c} 		&	\bcep	&	{\,}\tablefootmark{1} \\
$\nu$~Eri		&	229087301	&	04:36:19.14&	-03:21:08.86	&	3.74	&	3.93	&	B2\,III          		&	20		&				&	[5.62, 5.76]&	$<20$	&	Y \citep{Aerts1994c} 		&	\bcep	&	{\,}\tablefootmark{2} \\
$\eta$~Ori		&	4254645		&	05:24:28.62&	-02:23:49.73	&	3.18	&	3.35	&	B1\,V + B2:      		&	50\tablefootmark{3}	&	$\checkmark$		&	&	&	&	\bcep + EB	&	{\,}\tablefootmark{1} \\
$\eta$~Ori A		&	4254645		&	05:24:28.73&	-02:23:49.31	&	3.42	&	3.59	&	B1\,V           		&	130		&	$\checkmark$		&	7.69		&	$<1$		&	Y \citep{DeMey1996}	&	\bcep + EB	& \\ 
$\beta$~CMa		&	34590771		&	06:22:41.99&	-17:57:21.31	&	1.73	&	1.97	&	B1\,II-III       		&	$\simeq1$		&		&	[3.97, 4.18]&	$<11$	&	Y \citep{Aerts1994c} 		&	\bcep	&	{\,}\tablefootmark{4} \\
$\beta$~Cru		&	405567821	&	12:47:43.27&	-59:41:19.58	&	1.02	&	1.25	&	B1\,IV           		&	18		&	$\checkmark$	&	[5.23, 5.96]&	$<3$		&	Y \citep{Aerts1998c} 		&	\bcep	&	{\,}\tablefootmark{1, 2} \\
$\alpha$~Vir		&	178999156\tablefootmark{$\dagger$}	&	13:25:11.58&	-11:09:40.75	&	0.74	&	0.97	&	B1\,V		&	160		&	$\checkmark$	&	[0.50,7.50]&	$<4$		&	Y \citep{Smith_M_1985b} 	&	$-$	&	{\,}\tablefootmark{1, 5} \\ 
$\epsilon$~Cen		&	241398115	&	13:39:53.26&	-53:27:59.01	&	2.08	&	2.30	&	B1\,III          		&	115		&	?	&	[5.69, 7.65]&	$<1$		&	Y \citep{Schrijvers_PhD} 	&	\bcep	&	{\,}\tablefootmark{1, 6} \\
$\nu$~Cen		&	166447153	&	13:49:30.28&	-41:41:15.75	&	3.19	&	3.39	&	B2\,IV  			&	65		&	$\checkmark$	&	[2.35, 7.95]&	$<1$		&	Y \citep{Schrijvers2002b} 	&	\bcep + rot/bin	&	{\,}\tablefootmark{1} \\         
$\beta$~Cen		&	328329822	&	14:03:49.41&	-60:22:22.93	&	0.38	&	0.60	&	B1\,III			&	100		&	$\checkmark$	&	[6.41, 6.51]&	$<4$		&	Y \citep{Ausseloos2002b} 	&	\bcep	&	{\,}\tablefootmark{1, 2, 7} \\
$\alpha$~Lup		&	129117325	&	14:41:55.76&	-47:23:17.52	&	2.13	&	2.29	&	B1.5\,III       		&	$\simeq1$		&			&	[3.85, 4.22]&	$<7$		&	Y \citep{Mathias1994a} 	&	\bcep	&	{\,}\tablefootmark{2} \\
$\delta$~Lup		&	148415949	&	15:21:22.33&	-40:38:51.02	&	3.00	&	3.19	&	B1.5\,IV       		&	190		&	?			&	5.05		&	4		&	Y \citep{Lloyd1988} 		&	\bcep	&	{\,}\tablefootmark{1, 2} \\
$\epsilon$~Lup		&	147226597	&	15:22:40.87&	-44:41:22.61	&	3.19	&	3.37	&	B2\,IV-V       		&	40-170	&	$\checkmark$	&	10.36	&	$<1$		&	Y \citep{Uytterhoeven2005b}	&	\bcep	&	{\,}\tablefootmark{1, 6} \\
$\beta$~Sco		&	9921163\tablefootmark{$\dagger$}	&	16:05:26.23	&	-19:48:19.40	&	2.43	&	2.50	&	B1\,V + B2\,V	&	90	&	$\checkmark$	&	5.77		&	$<14$	&	Y \citep{Holmgren1997}	&	$-$	&	\\
$\beta^{1}$~Sco	&	9921163\tablefootmark{$\dagger$}	&	16:05:26.23	&	-19:48:19.63	&	2.55	&	2.62	&	B1\,V           	&	130	&	$\checkmark$	&	&	&	Y \citep{Holmgren1997} 	&	$-$	&	\\ 
$\omega^{1}$~Sco	&	161828468\tablefootmark{$\dagger$}&	16:06:48.43	&	-20:40:09.09	&	3.92	&	3.97	&	B1\,V           	&	110	&	&	$\simeq15$	&	15		&	Y \citep{Telting1998c} 	&	$-$	&	\\ 
$\sigma$~Sco		&	322928423\tablefootmark{$\dagger$}&	16:21:11.32	&	-25:35:34.05	&	3.02	&	2.89	&	O9.5\,V + B7\,V		&	50	&	$\checkmark$	&	[4.05, 4.17]	&	$<40$	&	Y \citep{Mathias1991c}	&	$-$	&	{\,}\tablefootmark{1, 2, 8} \\ 
$\theta$~Oph		&	83852015\tablefootmark{$\dagger$}	&	17:22:00.58&	-24:59:58.37	&	3.03	&	3.26	&	B2\,V + B5\,V		&	35		&	$\checkmark$	&	[7.12, 7.47]&	$<5$		&	Y \citep{Briquet2005} 	&	$-$	&	{\,}\tablefootmark{1, 2} \\
$\lambda$~Sco		&	465088681	&	17:33:36.52&	-37:06:13.76	&	1.48	&	1.62	&	B2\,IV + DA7.9		&	120			&	$\checkmark$	&	[3.57, 5.37]&	$<5$		&	Y \citep{Uytterhoeven2004b} 	&	\bcep + EB	&	{\,}\tablefootmark{1, 2, 9} \\
$\iota$~Her		&	164319891	&	17:39:27.89&	46:00:22.80	&	3.63	&	3.80	&	B3\,IV           		&	$\simeq0$		&	$\checkmark$	&	$\simeq8$	&	$<3$		&	Y \citep{Chapellier1987}	&	\bcep + SLF	&	{\,}\tablefootmark{1, 2} \\
$\kappa$~Sco		&	147868882	&	17:42:29.28&	-39:01:47.94	&	2.21	&	2.39	&	B1.5\,III        		&	115			&	$\checkmark$	&	[4.87, 5.00]&	$<2$		&	Y \citep{Uytterhoeven2005a}	&	\bcep	&	{\,}\tablefootmark{1, 2} \\
$\beta$~Cep		&	321818578	&	21:28:39.60&	70:33:38.57	&	3.01	&	3.23	&	B0.5\,III       		&	25			&	$\checkmark$	&	[4.92, 5.42]&	$<21$	&	Y \citep{Aerts1994b}		&	\bcep	&	{\,}\tablefootmark{1} \\

\hline
\multicolumn{14}{l}{Priority~1B: stars that are known to exhibit LPV from pulsations in the literature} \\
$\zeta$~CMa		&	124918771	&	06:20:18.79	&	-30:03:48.12	&	2.83	&	3.00	&	B2.5\,V          		&	25	&	$\checkmark$	&	&	&	Y \citep{Telting2006c}	&	SLF	&	{\,}\tablefootmark{1} \\
$\alpha$~Pyx		&	185815213	&	08:43:35.54	&	-33:11:10.99	&	3.50	&	3.68	&	B1.5\,III        		&	20	&	&	&	&	Y \citep{Telting2006c}	&	SLF	&	\\
$\delta$~Cru		&	390320354	&	12:15:08.72	&	-58:44:56.14	&	2.57	&	2.75	&	B2\,IV           		&	135	&	&	&	&	Y \citep{Telting2006c}	&	\bcep	&	\\ 
$\sigma$~Cen		&	21165132		&	12:28:02.38	&	-50:13:50.29	&	3.72	&	3.91	&	B2\,V            		&	170	&	&	&	&	Y \citep{Telting2006c}	&	\bcep	&	\\
$\alpha$~Mus		&	327150358	&	12:37:11.02	&	-69:08:08.03	&	2.48	&	2.65	&	B2\,IV           		&	115	&	?	&	&	&	Y \citep{Telting2006c}	&	SLF	&	{\,}\tablefootmark{1} \\
$\phi$~Cen		&	359762838	&	13:58:16.27	&	-42:06:02.71	&	3.61	&	3.80	&	B2\,IV          		&	80	&	&	&	&	Y \citep{Telting2006c}	&	\bcep	&	\\ 
$\upsilon^{1}$~Cen	&	359916090	&	13:58:40.75	&	-44:48:12.91	&	3.67	&	3.87	&	B2\,IV-V         		&	125	&	&	&	&	Y \citep{Telting2006c}	&	\bcep	&	\\
$\beta$~Lup		&	333947284	&	14:58:31.93	&	-43:08:02.27	&	2.46	&	2.68	&	B2\,III          		&	120	&	&	&	&	Y \citep{Schrijvers_PhD}	&	\bcep	&	\\
$\rho$~Sco		&	77541886		&	15:56:53.08	&	-29:12:50.66	&	3.66	&	3.86	&	B2\,IV-V         		&	100	&	$\checkmark$	&	&	&	Y \citep{Telting2006c}	&	\bcep	&	{\,}\tablefootmark{1} \\
$\kappa$~Cen		&	333950446	&	14:59:09.68	&	-42:06:15.11	&	2.93	&	3.13	&	B2\,IV           		&	30	&	?			&	&	&	Y \citep{Telting2006c}	&	\bcep	&	{\,}\tablefootmark{1, 10} \\
$\delta$~Sco		&	12725034\tablefootmark{$\dagger$}	&	16:00:20.01	&	-22:37:18.14	&	2.20	&	2.32	&	B0.3\,IV	&	150	&	$\checkmark$	&	&	&	Y \citep{Telting2006c}	&	$-$	&	{\,}\tablefootmark{1} \\
$\upsilon$~Sco		&	175627991	&	17:30:45.84	&	-37:17:44.93	&	2.47	&	2.70	&	B2\,IV           		&	30 	&	&	&	&	Y \citep{Telting2006c}	&	SLF	& 	{\,}\tablefootmark{11} \\

\hline
\multicolumn{14}{l}{Priority~1C: stars that exhibit high-frequency pulsation modes in TESS photometry but lack spectroscopic confirmation of LPV in the literature} \\
$\zeta$~Cas		&	240669906	&	00:36:58.28	&	53:53:48.87	&	3.47	&	3.66	&	B2\,IV			&	10\tablefootmark{12}		&	&	&	&	N \citep{Telting2006c}	&	\bcep	&	{\,}\tablefootmark{13} \\  
a~Car			&	385220745	&	09:10:58.09	&	-58:58:00.82	&	3.25	&	3.40	&	B2\,IV-V         		&	30	&	$\checkmark$	&	&	&	N \citep{Telting2006c}	&	\bcep + SLF	&	{\,}\tablefootmark{1, 10} \\ 
$\rho$~Cen		&	334217681	&	12:11:39.13	&	-52:22:06.41	&	3.81	&	3.96	&	B3\,V           		&	140\tablefootmark{3}		&	?	&	&	&	&	\bcep	&	{\,}\tablefootmark{1, 14} \\
$\alpha^{1}$~Cru	&	450568754	&	12:26:35.90	&	-63:05:56.73	&	1.10	&	1.28	&	B0.5\,IV         		&	120\tablefootmark{3}		&	$\checkmark$	&	&	&	&	\bcep	&	{\,}\tablefootmark{1, 15} \\
$\alpha^{2}$~Cru	&	450568754	&	12:26:36.44	&	-63:05:58.28	&	1.41	&	1.58	&	B1\,V            		&	200\tablefootmark{3}		&	$\checkmark$	&	&	&	&	\bcep	&	{\,}\tablefootmark{1, 15} \\
$\gamma$~Mus	&	360263464	&	12:32:28.01	&	-72:07:58.76	&	3.73 	&	3.88	&	B3\,V            		&	190\tablefootmark{3}		&	&	&	&	&	\bcep/SPB&	{\,}\tablefootmark{16, 17} \\
$\beta$~Mus		&	329162201	&	12:46:16.80	&	-68:06:29.22	&	3.38	&	3.04	&	B2\,V + B3\,V  		&	140\tablefootmark{18}	&	$\checkmark$	&	&	&	&	\bcep	&	{\,}\tablefootmark{1, 14, 19} \\
$\eta$~Lup		&	59095516		&	16:00:07.33	&	-38:23:48.15	&	3.19	&	3.41	&	B2.5\,IV         		&	190\tablefootmark{18}	&	?	&	&	&	&	\bcep	&	{\,}\tablefootmark{1, 20} \\

\hline
\multicolumn{14}{l}{Priority~2: stars with high-amplitude g-mode pulsations (found in TESS light curves and LPV studies and/or photometry in literature)} \\
$\gamma$~Cas	&	51962733		&	00:56:42.53	&	60:43:00.27	&	2.29	&	2.39	&	B0.5\,IV\,pe 		&	295\tablefootmark{12}	&	$\checkmark$	&	&	&	Y \citep{Yang_S_1988}	&	SPB		&	{\,}\tablefootmark{1, 21} \\
$\kappa$~CMa		&	52982382		&	06:49:50.46	&	-32:30:30.52	&	3.68	&	3.89	&	B1.5\,Ve         		&	220\tablefootmark{22}	&	&	&	&	Y \citep{Rivinius2003b}	&	SPB		&	{\,}\tablefootmark{21} \\
$\omega$~CMa	&	65903024		&	07:14:48.65	&	-26:46:21.61	&	3.64	&	3.82	&	B2.5\,V\,e        		&	80\tablefootmark{22}		&	&	&	&	Y \citep{Stefl2003b}		&	SPB		&	{\,}\tablefootmark{21} \\
omi~Vel  			&	93549165		&	08:40:17.59	&	-52:55:18.80	&	3.44	&	3.63	&	B3/5(V)			&	40	&	?	&	&	&	Y \citep{Aerts1999a} 	&	SPB		&	{\,}\tablefootmark{1, 2} \\
$\delta$~Cen		&	333670784	&	12:08:21.50	&	-50:43:20.74	&	2.39	&	2.52	&	B2\,V\,ne          		&	220\tablefootmark{22}	&	?	&	&	&	Y \citep{Rivinius2003b}	&	SPB		&	{\,}\tablefootmark{1, 14, 21} \\
$\eta$~UMa		&	219033887	&	13:47:32.44	&	49:18:47.76	&	1.67	&	1.86	&	B3\,V            		&	150\tablefootmark{12}	&	&	&	&	N \citep{Smith_M_2006d}	&	SPB		&	\\
$\mu$~Cen		&	243647020	&	13:49:36.99	&	-42:28:25.43	&	3.27	&	3.43	&	B2\,V\,npe         	&	155\tablefootmark{22}	&	?	&	&	&	Y \citep{Rivinius2001b}	&	SPB		&	{\,}\tablefootmark{1, 14, 21} \\
$\eta$~Cen		&	128116539	&	14:35:30.42	&	-42:09:28.17	&	2.12	&	2.31	&	B2\,V\,e           		&	350	&	&	&	&	Y \citep{Janot-Pacheco1999} 	&	SPB	&	{\,}\tablefootmark{23} \\
$\zeta$~Oph		&	152859121\tablefootmark{$\dagger$}&	16:37:09.54	&	-10:34:01.53	&	2.58	&	2.56	&	O9.2\,IV\,nn	&	320		&	&	&	&	Y \citep{Kambe1997} 	&	$-$	&	{\,}\tablefootmark{2, 21} \\
$\alpha$~Ara		&	412724758	&	17:31:50.49	&	-49:52:34.12	&	2.78	&	2.95	&	B2\,V\,ne          		&	250\tablefootmark{22}	&	&	&	&	Y \citep{Rivinius2003b}	&	SPB	&	{\,}\tablefootmark{21} \\
$\alpha$~Tel		&	90482097		&	18:26:58.42	&	-45:58:06.45	&	3.30	&	3.46	&	B3\,IV          		&	35	&	&	&	&	Y \citep{Telting2006c}	&	SPB	&	{\,}\tablefootmark{1} \\

\hline
\multicolumn{14}{l}{Priority~3: stars dominated by SLF in TESS photometry} \\
$\epsilon$~Cas		&	10510382		&	01:54:23.73	&	63:40:12.37	&	3.22	&	3.37	&	B3\,Vp\,sh			&	30\tablefootmark{12}		&	&	&	&	&	SLF	&	{\,}\tablefootmark{17} \\
$\zeta$~Per		&	94367286		&	03:54:07.92	&	31:53:01.08	&	2.97	&	2.85	&	B1\,Ib          		&	40\tablefootmark{12}		&	?	&	&	$\simeq3$&	Y \citep{Simon-Diaz2010b}	&	SLF	& 	{\,}\tablefootmark{1} \\
$\pi^{4}$~Ori		&	452810783	&	04:51:12.36	&	05:36:18.37	&	3.50	&	3.68	&	B2\,III+B2\,IV      	&	35\tablefootmark{12}		&	$\checkmark$	&	&	&	Y \citep{Telting2006c}	&	SLF	&	{\,}\tablefootmark{1, 17} \\
$\gamma$~Ori		&	365572007	&	05:25:07.86	&	06:20:58.93	&	1.42	&	1.64	&	B2\,V            		&	55	&	&	&	&	N \citep{Telting2006c}	&	SLF	&	{\,}\tablefootmark{4, 24} \\ 
$\zeta$~Ori B		&	712837934	&	05:40:45.57	&	-01:56:35.59	&	3.50	&	3.70	&	O9.5\,II-III\,(n)  		&	&	$\checkmark$	&	&	&	&	SLF	&	\tablefootmark{1, 25} \\ 
$\iota$~Ori		&	427395774	&	05:35:25.98	&	-05:54:35.64	&	2.53	&	2.77	&	O9\,III\,var       		&	130\tablefootmark{3}		&	$\checkmark$	&	&	&	&	SLF	&	{\,}\tablefootmark{1, 4, 26} \\
$\epsilon$~Ori		&	427451176	&	05:36:12.81	&	-01:12:06.91	&	1.51	&	1.69	&	B0\,Ia           		&	65\tablefootmark{12}		&	&	&	$\simeq13$	&	Y \citep{Simon-Diaz2010b}	&	SLF	&	{\,}\tablefootmark{4, 27} \\ 
$\zeta$~Ori		&	11360636		&	05:40:45.53	&	-01:56:33.26	&	1.59	&	1.79	&	O9.7\,Ib + B0\,III   	&	&	$\checkmark$	&	&	&	&	SLF	&	\\
$\zeta$~Ori A		&	11360636		&	05:40:45.53	&	-01:56:33.26	&	1.77	&	1.88	&	O9.2\,Ib\,varNwk   	&	140\tablefootmark{3}		&	$\checkmark$	&	&	$\simeq12$	&	Y \citep{Simon-Diaz2010b}	&	SLF	&	{\,}\tablefootmark{25} \\ 
$\kappa$~Ori		&	66651575		&	05:47:45.39	&	-09:40:10.58	&	1.88	&	2.06	&	B0.5\,Ia         		&	65\tablefootmark{12}		&	&	&	$\simeq10$&	Y \citep{Simon-Diaz2010b}	&	SLF	&	{\,}\tablefootmark{4, 28} \\
$\epsilon$~CMa	&	63198307		&	06:58:37.55	&	-28:58:19.51	&	1.29	&	1.50	&	B1.5\,II         		&	25\tablefootmark{12}		&	?	&	&	&	&	SLF	&	{\,}\tablefootmark{1, 4, 28} \\ 
omi$^{2}$~CMa	&	80466973		&	07:03:01.47	&	-23:49:59.85	&	2.94	&	3.02	&	B3\,Ia          		&	35\tablefootmark{12}		&	&	&	&	&	SLF	&	{\,}\tablefootmark{2, 4} \\
$\chi$~Car		&	269407223	&	07:56:46.72	&	-52:58:56.39	&	3.27	&	3.43	&	B3\,IV  			&	95\tablefootmark{3}		&	&	&	&	N \citep{Leone1998b}	&	SLF	&	{\,}\tablefootmark{1, 29} \\         
$\kappa$~Vel		&	387106852	&	09:22:06.82	&	-55:00:38.40	&	2.32	&	2.47	&	B2\,IV  			&	50\tablefootmark{12}		&	$\checkmark$	&	&	&	&	SLF	&	{\,}\tablefootmark{1, 10, 14, 15} \\    
$\iota$~Lup		&	242497929	&	14:19:24.22	&	-46:03:29.14	&	3.36	&	3.53	&	B2.5\,IV			&	370\tablefootmark{3}		&	&	&	&	&	SLF	&	{\,}\tablefootmark{30} \\
$\tau$~Sco		&	205175750	&	16:35:52.95	&	-28:12:57.66	&	2.56	&	2.81	&	B0.2\,V          		&	0\tablefootmark{12}		&	&	&	&	N \citep{Telting2006c} &	SLF	&	\\ 
$\gamma$~Ara		&	325709821	&	17:25:23.66	&	-56:22:39.81	&	3.21	&	3.34	&	B1\,Ib           		&	280\tablefootmark{3}		&	?	&	&	&	Y \citep{Baade1983c} &	SLF	&	{\,}\tablefootmark{1} \\ 
$\theta$~Ara		&	364875523	&	18:06:37.87	&	-50:05:29.31	&	3.58	&	3.66	&	B2\,Ib           		&	120\tablefootmark{3}		&	&	&	&	Y \citep{Baade1983c} &	SLF	&	\\ 
$\alpha$~Pav		&	219974785	&	20:25:38.86	&	-56:44:06.32	&	1.79	&	1.92	&	B2\,IV           		&	40	&	$\checkmark$	&	&	&	Y \citep{Telting2006c}	&	SLF	&	{\,}\tablefootmark{1, 15} \\

\hline
\multicolumn{14}{l}{No evidence of \bcep pulsations in the literature or in TESS light curves (but other variability may be present)} \\
omi~Per			&	101255849	&	03:44:19.13	&	32:17:17.69	&	3.88	&	3.83	&	B1\,III        		&	90\tablefootmark{12}		&	$\checkmark$	&	&	&	N \citep{Telting2006c}	&	rot/bin	&	{\,}\tablefootmark{1, 23, 31} \\
$\pi^{5}$~Ori		&	262435444	&	04:54:15.10	&	02:26:26.41	&	3.54	&	3.73	&	B2\,III          		&	90\tablefootmark{12}		&	$\checkmark$	&	&	&	&	rot/bin	&	{\,}\tablefootmark{1, 23, 31} \\
$\eta$~Aur		&	122385146\tablefootmark{$\ddagger$}	&	05:06:30.89	&	41:14:04.11	&	3.00	&	3.18	&	B3\,V            		&	95\tablefootmark{12}		&	&	&	&	&	$-$		&	{\,}\tablefootmark{17, 32} \\
$\delta$~Ori		&	50743469		&	05:32:00.40	&	-00:17:56.74	&	2.02	&	2.41	&	B0\,III + O9\,V		&	130\tablefootmark{12}	&	$\checkmark$	&	&	&	Y \citep{Kholtygin2007d}	&	EB+SLF	&	{\,}\tablefootmark{1, 4} \\
$\zeta$~Tau		&	19704573\tablefootmark{$\dagger$}	&	05:37:38.69	&	21:08:33.16	&	2.84	&	3.03	&	B1\,IV\,e\,sh		&	125\tablefootmark{12}	&	$\checkmark$	&	&	&	Y \citep{Rivinius2003b}	&	$-$	&	{\,} \tablefootmark{1, 21} \\
$\sigma$~Ori		&	11286198		&	05:38:44.77	&	-02:36:00.28	&	3.58	&	3.79	&	O9.5\,V          		&	95\tablefootmark{12}		&	$\checkmark$	&	&	&	N \citep{Telting2006c}	&	EB	&	{\,}\tablefootmark{4, 33} \\
$\epsilon$~Car B	&	342884451\tablefootmark{$\ddagger$}	&	08:22:30.88	&	-59:30:34.40	&	3.93	&	3.85	&	B2\,V\,p           	&	&	&	&	&	&	$-$	&	\\
i~Car			&	357639411\tablefootmark{$\ddagger$}	&	09:11:16.72	&	-62:19:01.13	&	3.77	&	3.94	&	B3\,V            	&	20\tablefootmark{3}	&	&	&	&	N \citep{Telting2006c}	&	$-$	&	{\,}\tablefootmark{17} \\ 
$\rho$~Leo		&	392834850\tablefootmark{$\dagger$}&	10:32:48.67	&	09:18:23.71	&	3.72	&	3.87	&	B1\,Iab        		&	50\tablefootmark{12}		&	?	&	&	&	N \citep{Aerts2018a}	&	$-$	&	{\,}\tablefootmark{1} \\
$\theta$~Car		&	390442076	&	10:42:57.40	&	-64:23:40.02	&	2.54	&	2.76	&	B0\,V\,p           		&	150\tablefootmark{12} 	&	$\checkmark$	&	&	&	N \citep{Telting2006c}	&	SLF	&	{\,}\tablefootmark{1, 34} \\
$\zeta$~Cen		&	113350416	&	13:55:32.39	&	-47:17:18.15	&	2.33	&	2.55	&	B2.5\,IV         		&	220\tablefootmark{3}		&	$\checkmark$	&	&	&	&	rot/bin	&	{\,}\tablefootmark{1, 15} \\
$\gamma$~Lup		&	171611328	&	15:35:08.45	&	-41:10:00.32	&	2.59	&	2.77	&	B2\,IV           		&	270\tablefootmark{18}	&	$\checkmark$	&	&	&	&	rot/bin	&	{\,}\tablefootmark{1, 2, 15, 19} \\
$\tau$~Lib		&	147883533	&	15:38:39.37	&	-29:46:39.90	&	3.48	&	3.64	&	B2.5\,V			&	135	&	$\checkmark$	&	&	&	N \citep{Telting2006c}	&	rot/bin	&	{\,}\tablefootmark{1, 30} \\
$\pi$~Sco			&	66744681\tablefootmark{$\dagger$}	&	15:58:51.11	&	-26:06:50.79	&	2.71	&	2.91	&	B1\,V + B2:\,V:	&	100	&	$\checkmark$	&	&	&	? \citep{Telting2006c}	&	EB	&	{\,}\tablefootmark{1, 23} \\
$\mu^{1}$~Sco		&	338104678	&	16:51:52.22	&	-38:02:50.64	&	2.82	&	2.98	&	B1\,V + B          	&	240\tablefootmark{3}		&	$\checkmark$	&	&	&	&	EB	& {\,}\tablefootmark{1, 23} \\
$\mu^{2}$~Sco		&	338105219	&	16:52:20.15	&	-38:01:03.13	&	3.34	&	3.54	&	B2\,IV          		&	50	&	&	&	&	Y \citep{Telting2006c}	&	EB	&	{\,}\tablefootmark{11} \\
$\sigma$~Sgr		&	91093307\tablefootmark{$\dagger$}	&	18:55:15.93	&	-26:17:48.21	&	1.92	&	2.07	&	B2\,V	&	165	&	&	&	&	N \citep{Telting2006c}	&	$-$	&	\\

\end{longtable} 

\tablefoot{Columns are star name, TESS input catalogue (TIC) ID number, coordinates, $B$ and $V$ magnitudes, spectral type as listed by {\sc simbad} and projected rotation velocity, $v\,\sin\,i$ (if known). In the `bin' column, a `$\checkmark$' symbol corresponds to if a binary or higher-order multiple system orbital solution has been determined, whereas a `$?$' symbol means evidence for binarity has been found. The approximate frequency range of known \bcep pulsation modes and the corresponding radial velocities measured using previous LPV studies are also provided if available. A reference is provided for a dedicated search for LPVs caused by pulsations, which if successful is indicated by `Y', unsuccessful by `N' and inconclusive by `?'. We also provide a variability classification based on our analysis of TESS light curves.}
\tablefoottext{$\dagger$}{Not observed by TESS to date,}
\tablefoottext{$\ddagger$}{Problematic light curve extraction and/or dominated by systematics,}
\tablefoottext{1}{Evidence of multiplicity from \citet{Eggleton2008e},} 
\tablefoottext{2}{Stars known to have \bcep or SPB pulsations from Hipparcos photometry by \citet{Lefevre2009c},} 
\tablefoottext{3}{Projected rotational velocity from \citet{Hoffleit1991},} 
\tablefoottext{4}{Detection of photometric variability using TESS data by \citet{Burssens2020a},} 
\tablefoottext{5}{Interferometric masses determined by \citet{Herbison-Evans1971},} 
\tablefoottext{6}{Detection of photometric variability by \citet{Shobbrook1972c},} 
\tablefoottext{7}{Interferometric masses determined by \citet{Davis_J_2005} and \citet{Ausseloos2006},} 
\tablefoottext{8}{Interferometric masses determined by \citet{North_J_2007b},} 
\tablefoottext{9}{Interferometric masses determined by \citet{Tango2006},} 
\tablefoottext{10}{Detection of photometric variability by \citet{Daszy2017b},} 
\tablefoottext{11}{Classified as a constant photometric star by \citet{Handler2013c},} 
\tablefoottext{12}{Projected rotational velocity from \citet{Abt2002},} 
\tablefoottext{13}{Known magnetic SPB star \citet{Neiner2003a, Briquet2007a},} 
\tablefoottext{14}{Detection of photometric variability by \citet{Jakate1978d},} 
\tablefoottext{15}{Non-detection of photometric variability by \citet{Jerzykiewicz1977b},}
\tablefoottext{16}{Identified as an SPB star from Hipparcos photometry by \citet{Aerts1999a},} 
\tablefoottext{17}{Identified as candidate pulsating star using Hipparcos photometry by \citet{Koen2002c},} 
\tablefoottext{18}{Projected rotational velocity from \citet{Wolff2007},}
\tablefoottext{19}{Non-detection of photometric variability by \citet{Jakate1979d},} 
\tablefoottext{20}{Non-detection of photometric variability by \citet{Shobbrook1978b},} 
\tablefoottext{21}{(Pulsating) Be star discussed in review paper by \citet{Rivinius2003b},} 
\tablefoottext{22}{Projected rotational velocity from \citet{Rivinius2003b},} 
\tablefoottext{23}{Identified as (possible) eclipsing binary using Hipparcos photometry by \citet{Lefevre2009c},} 
\tablefoottext{24}{Detection of photometric variability by \citet{Krisciunas1994b},} 
\tablefoottext{25}{Detection of photometric variability by \citet{Buysschaert2017a},} 
\tablefoottext{26}{Detection of photometric variability using BRITE data by \citet{Pablo2017a},} 
\tablefoottext{27}{Detection of LPV also by \citet{Prinja2004a},} 
\tablefoottext{28}{Spectroscopic modelling performed by \citet{Haucke2018a},} 
\tablefoottext{29}{Detection of photometric variability by \citet{Elst1979c},} 
\tablefoottext{30}{Non-detection of photometric variability by \citet{Sterken1983b},}
\tablefoottext{31}{Detection of photometric variability by \citet{Waelkens1983h},} 
\tablefoottext{32}{Detection of photometric variability using BRITE data by \citet{Strassmeier2020},}
\tablefoottext{33}{Spectroscopic modelling performed by \citet{Simon-Diaz2015b},} 
\tablefoottext{34}{Identified as candidate pulsating star using Hipparcos photometry by \citet{Adelman2001d}} 

\end{landscape}
}

\onecolumn
\normalsize

\newpage

\section{TESS light curves of CubeSpec priority~1A targets}
\label{section: appendix: priority1}

\begin{figure*}
\centering
\includegraphics[width=0.433\textwidth]{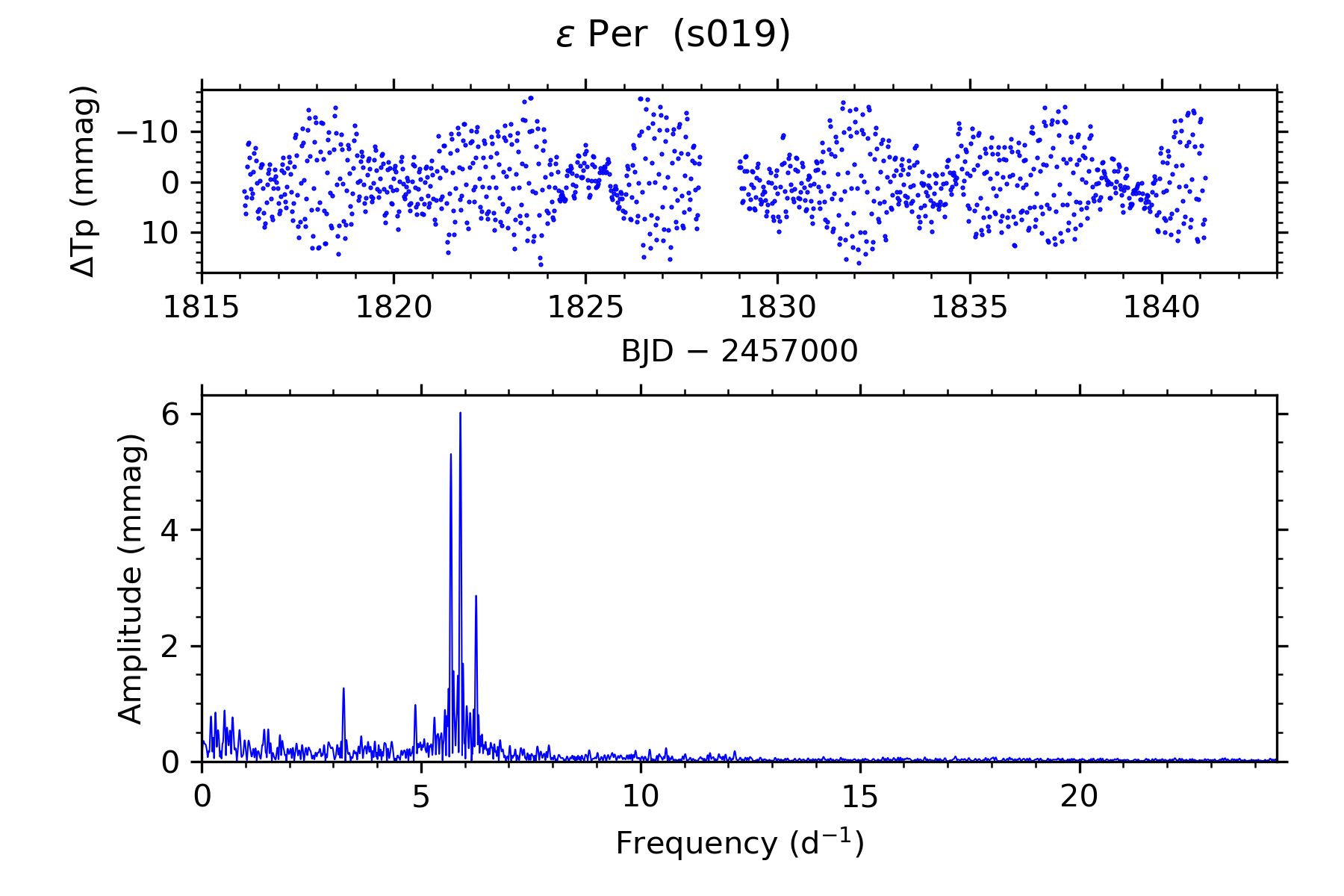}
\includegraphics[width=0.433\textwidth]{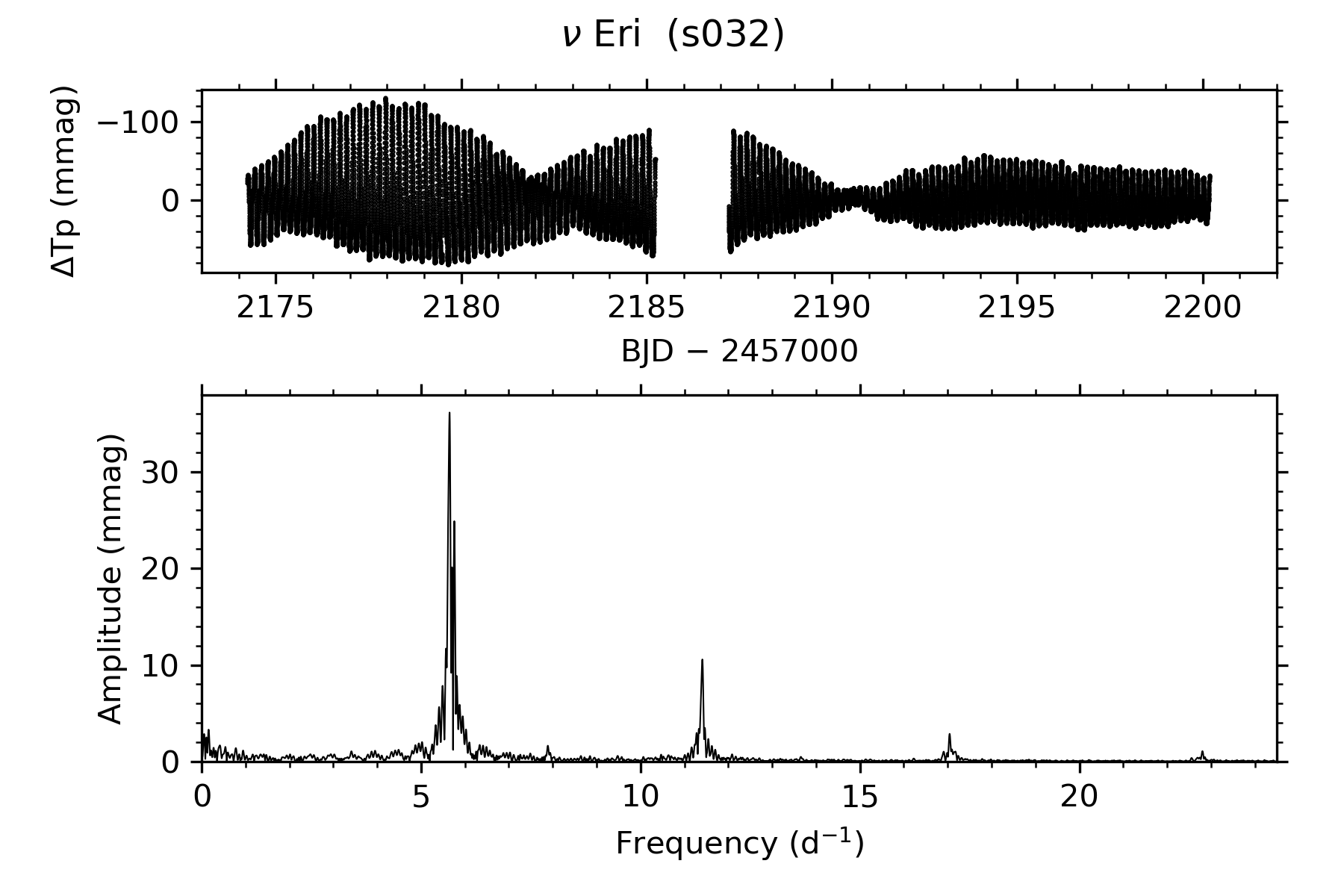}
\includegraphics[width=0.433\textwidth]{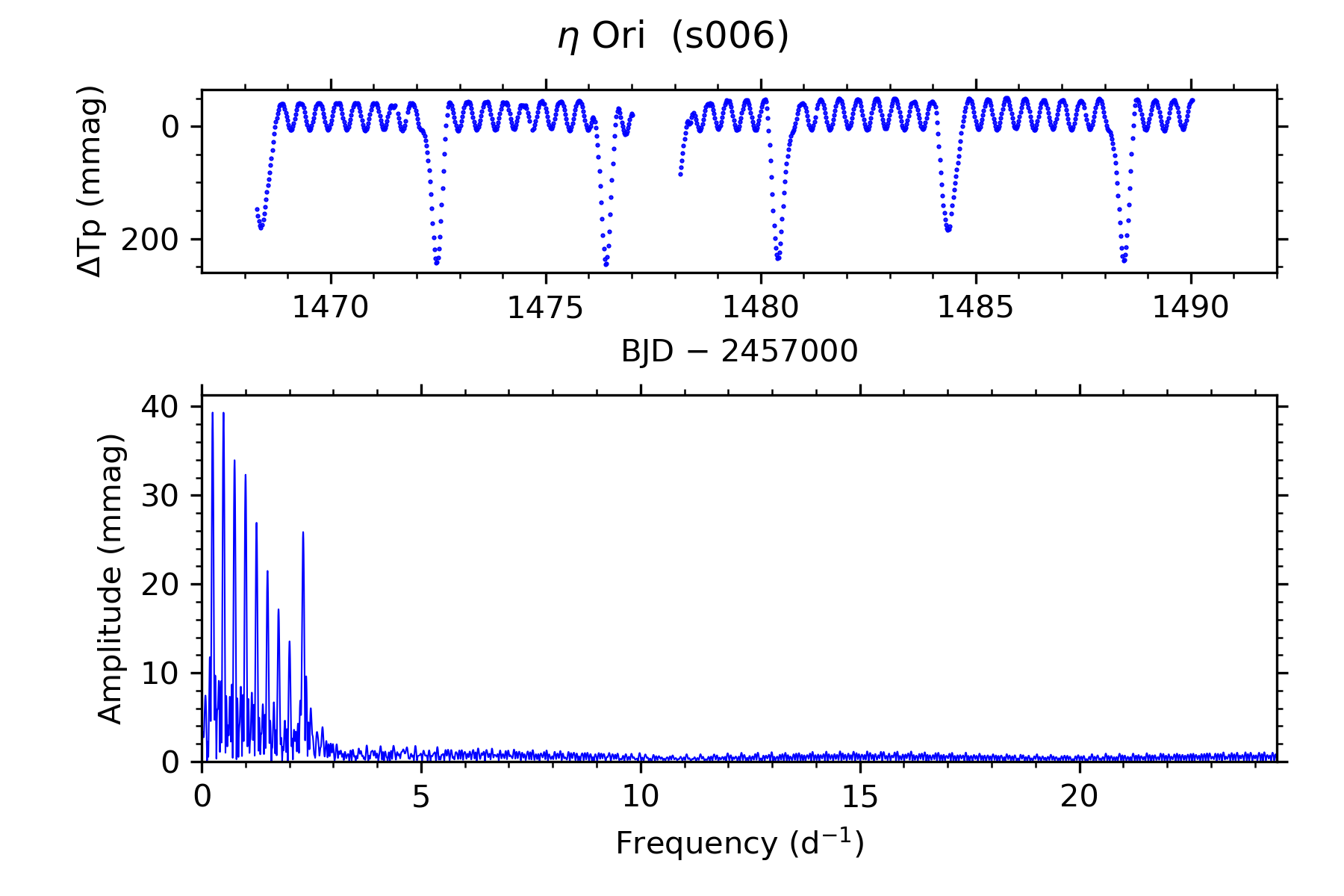}
\includegraphics[width=0.433\textwidth]{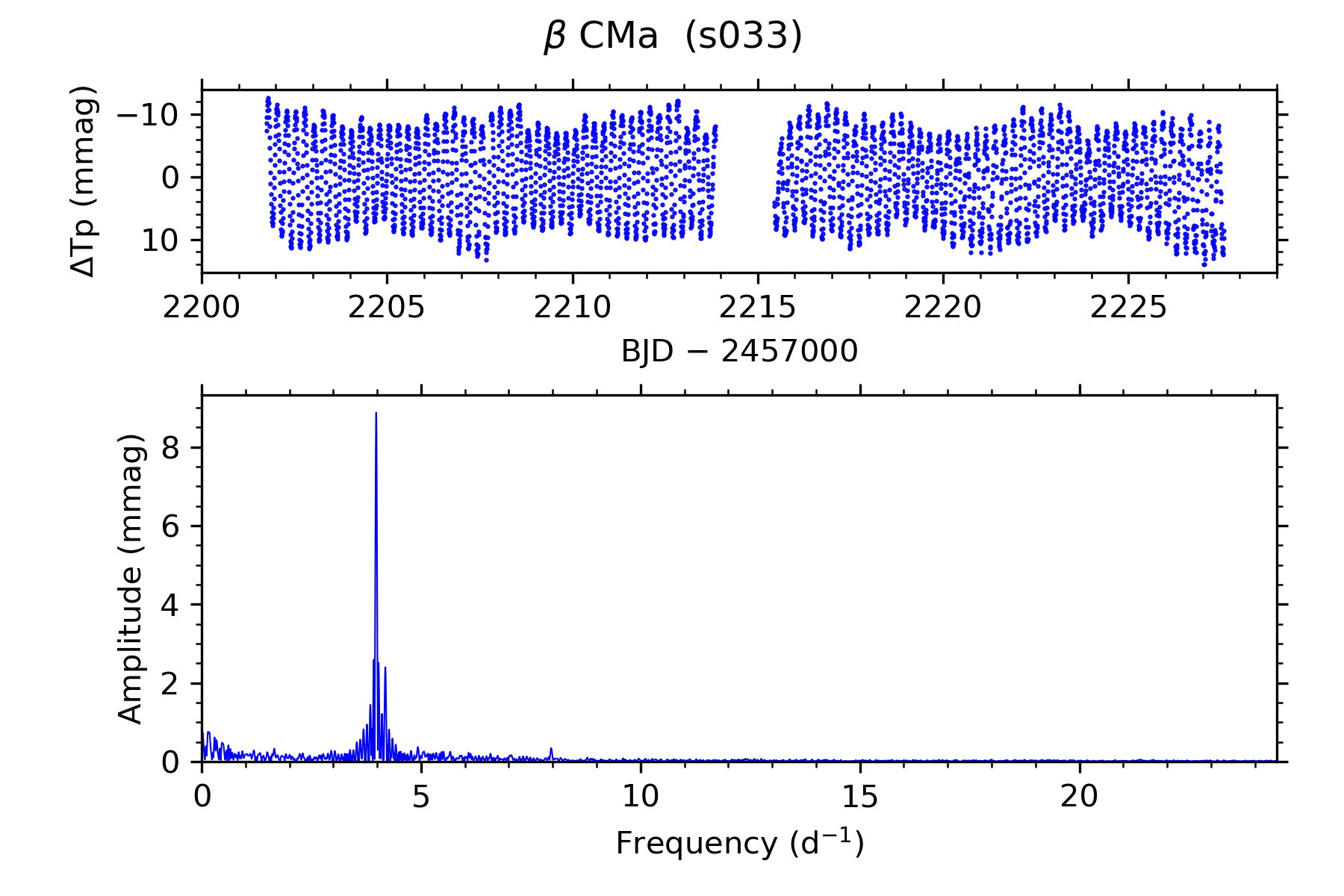}
\includegraphics[width=0.433\textwidth]{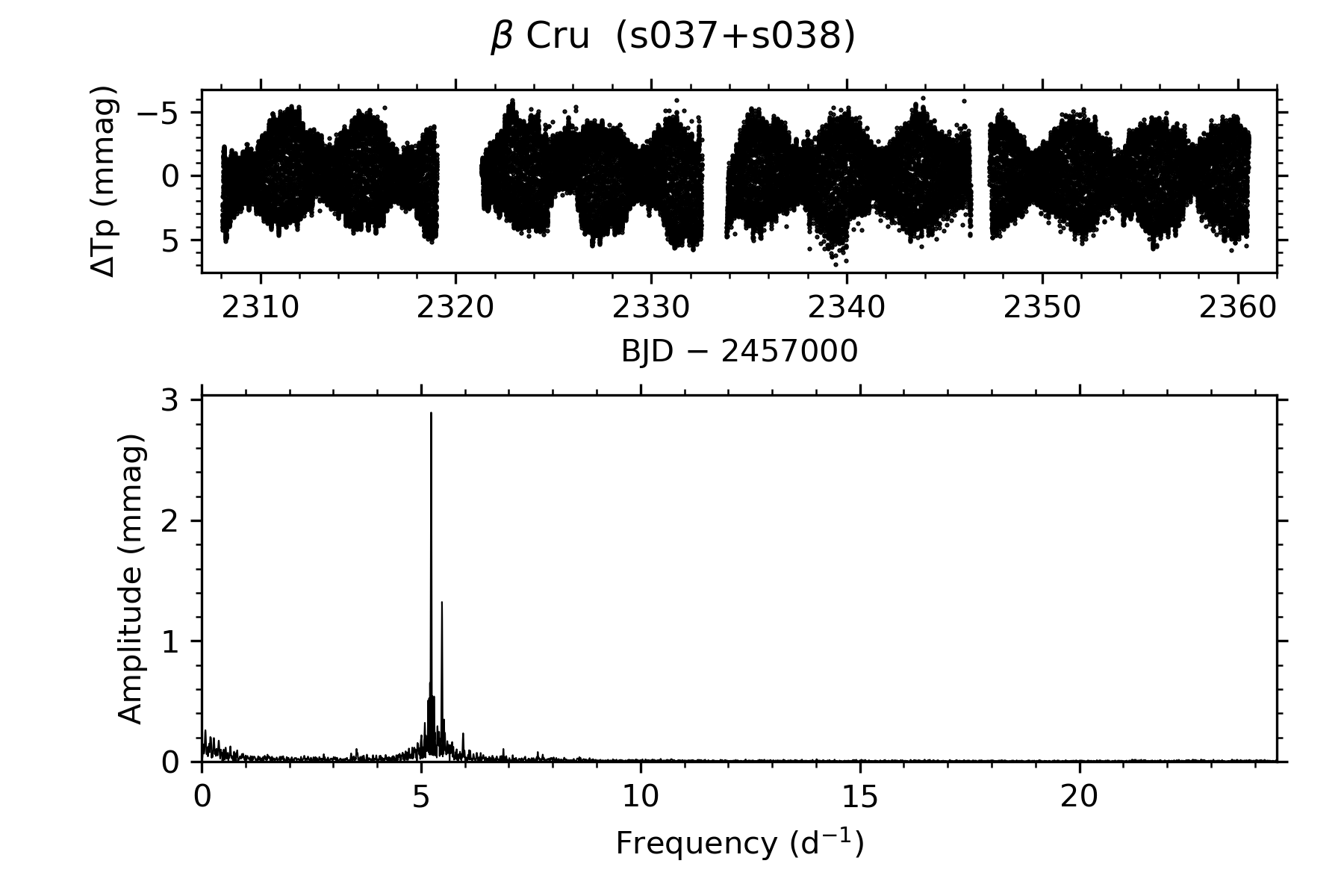}
\includegraphics[width=0.433\textwidth]{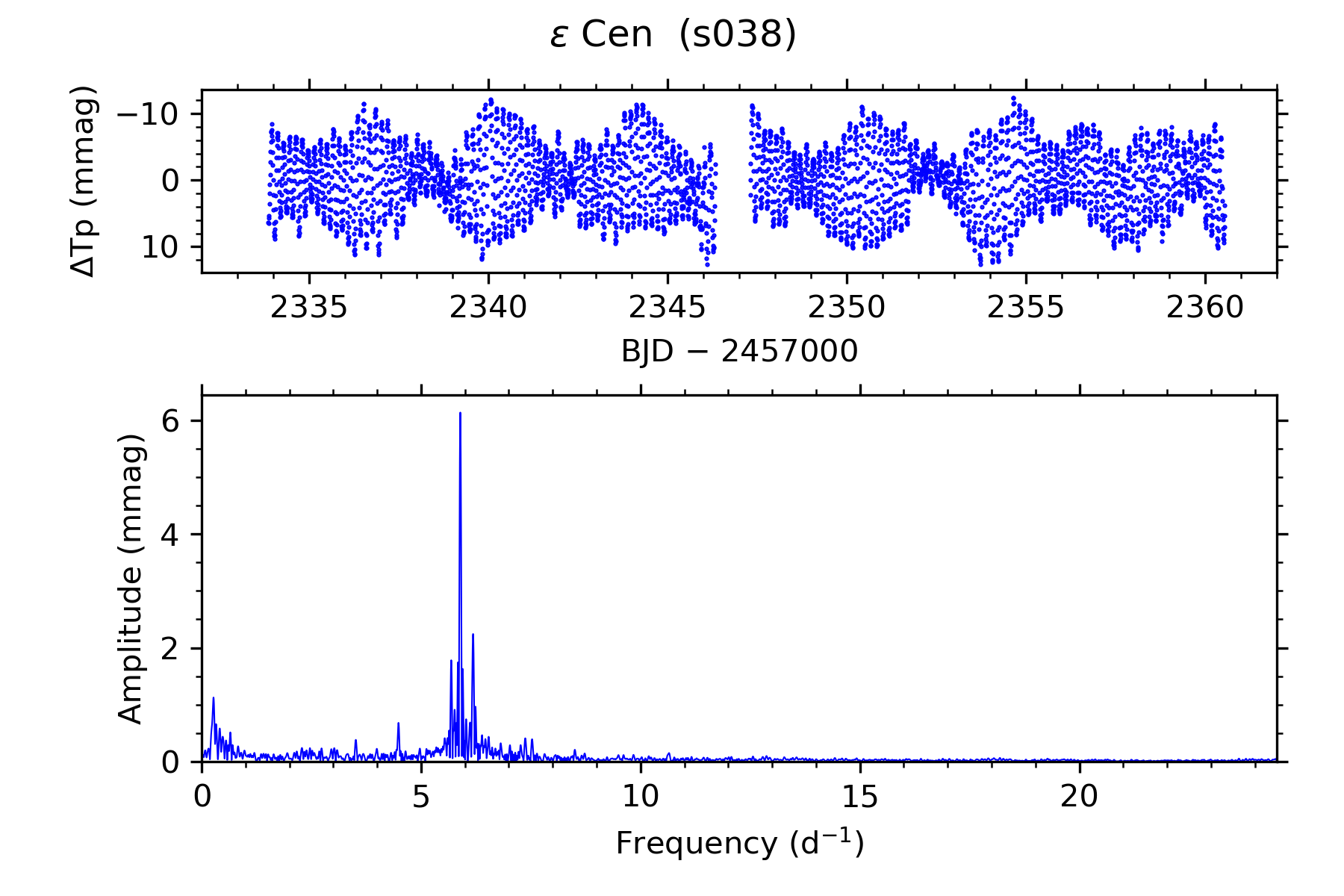}
\includegraphics[width=0.433\textwidth]{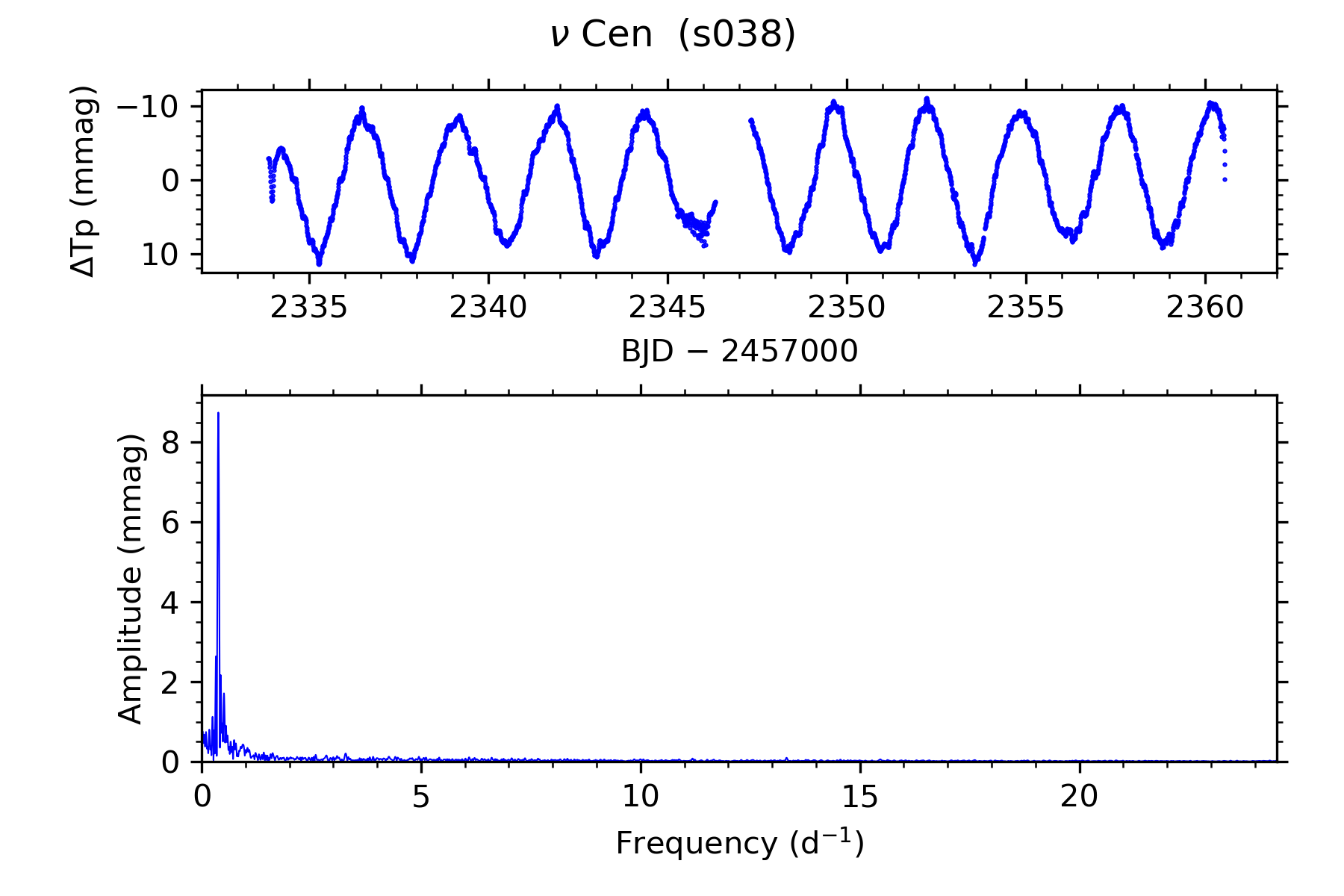}
\includegraphics[width=0.433\textwidth]{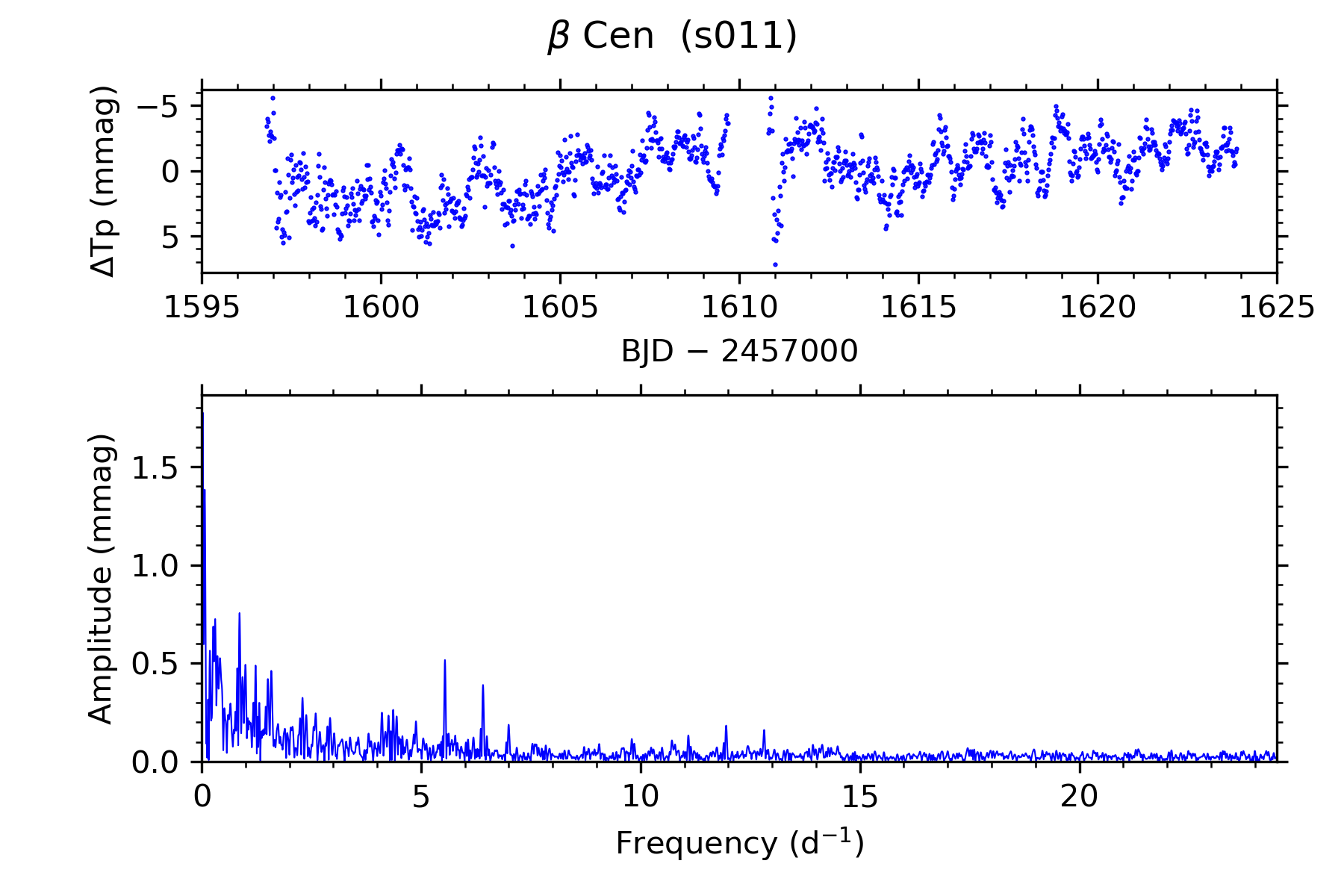}
\caption{TESS light curves and amplitude spectra for priority~1A CubeSpec mission targets. Those shown in black are 2-min cadence PDC-SAP light curves from the SPOC pipeline. Those in blue are light curves extracted using our own simple aperture photometry tools (see text for details).}
\label{figure: priority1_1}
\end{figure*}

\begin{figure*}
\centering
\includegraphics[width=0.45\textwidth]{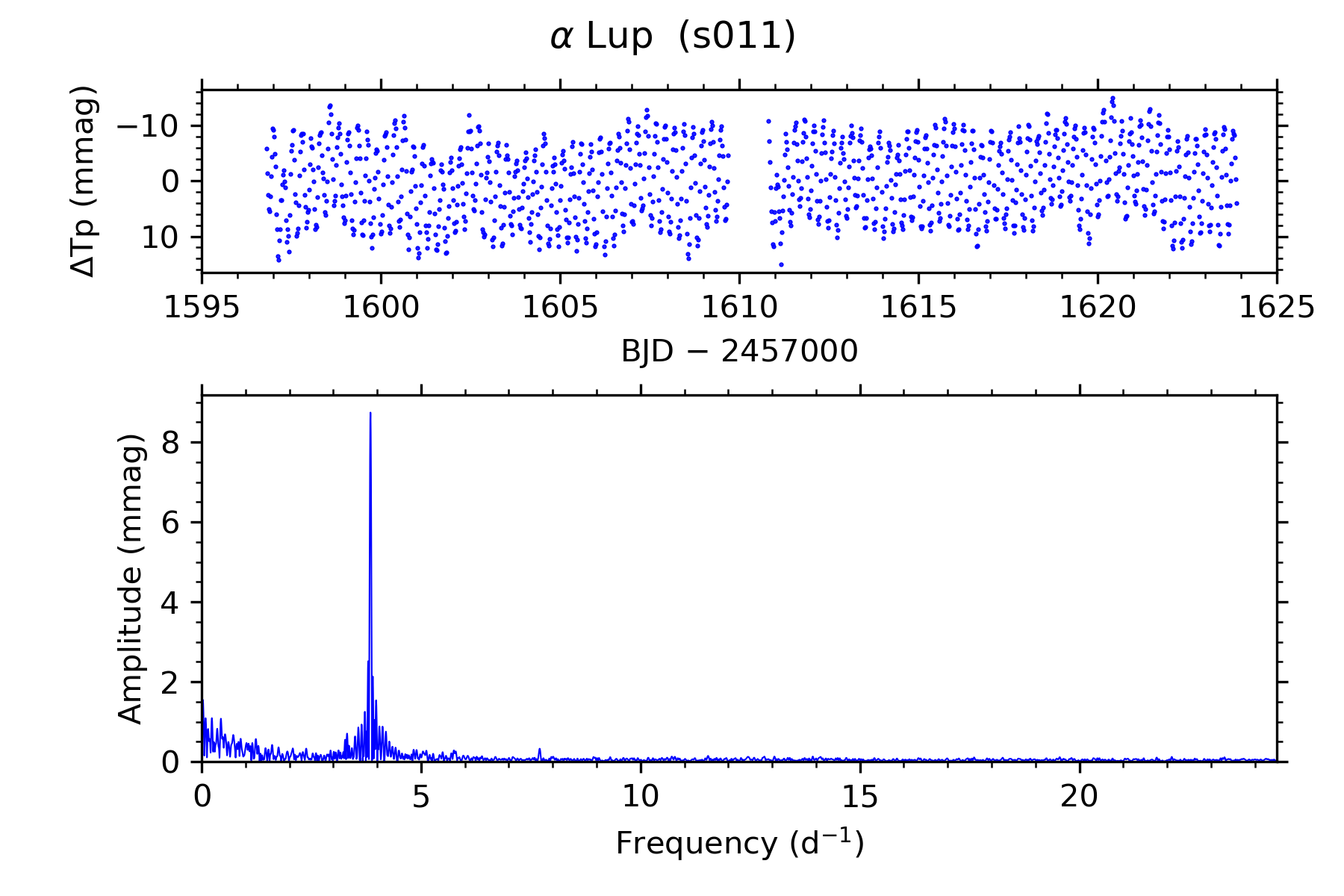}
\includegraphics[width=0.45\textwidth]{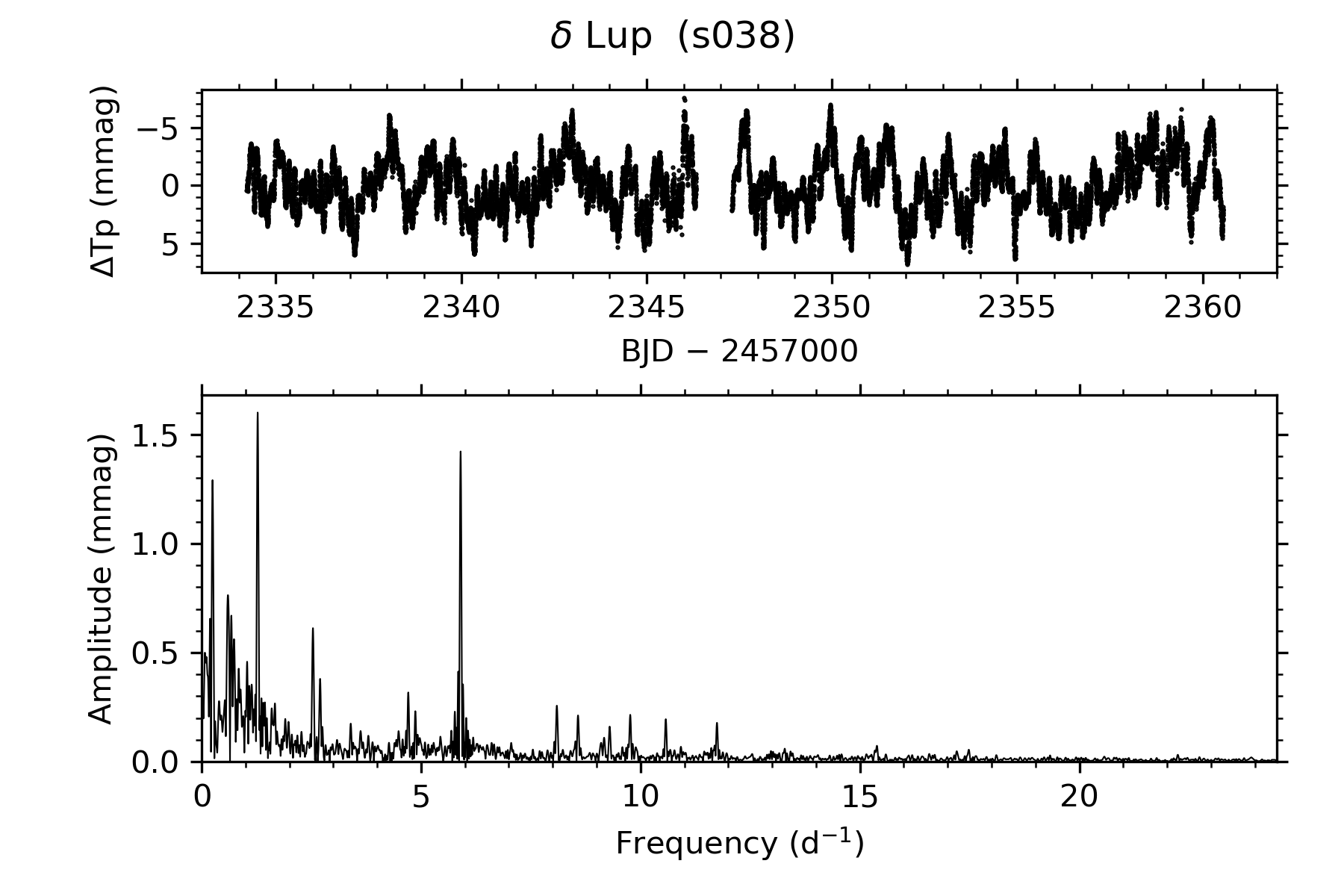}
\includegraphics[width=0.45\textwidth]{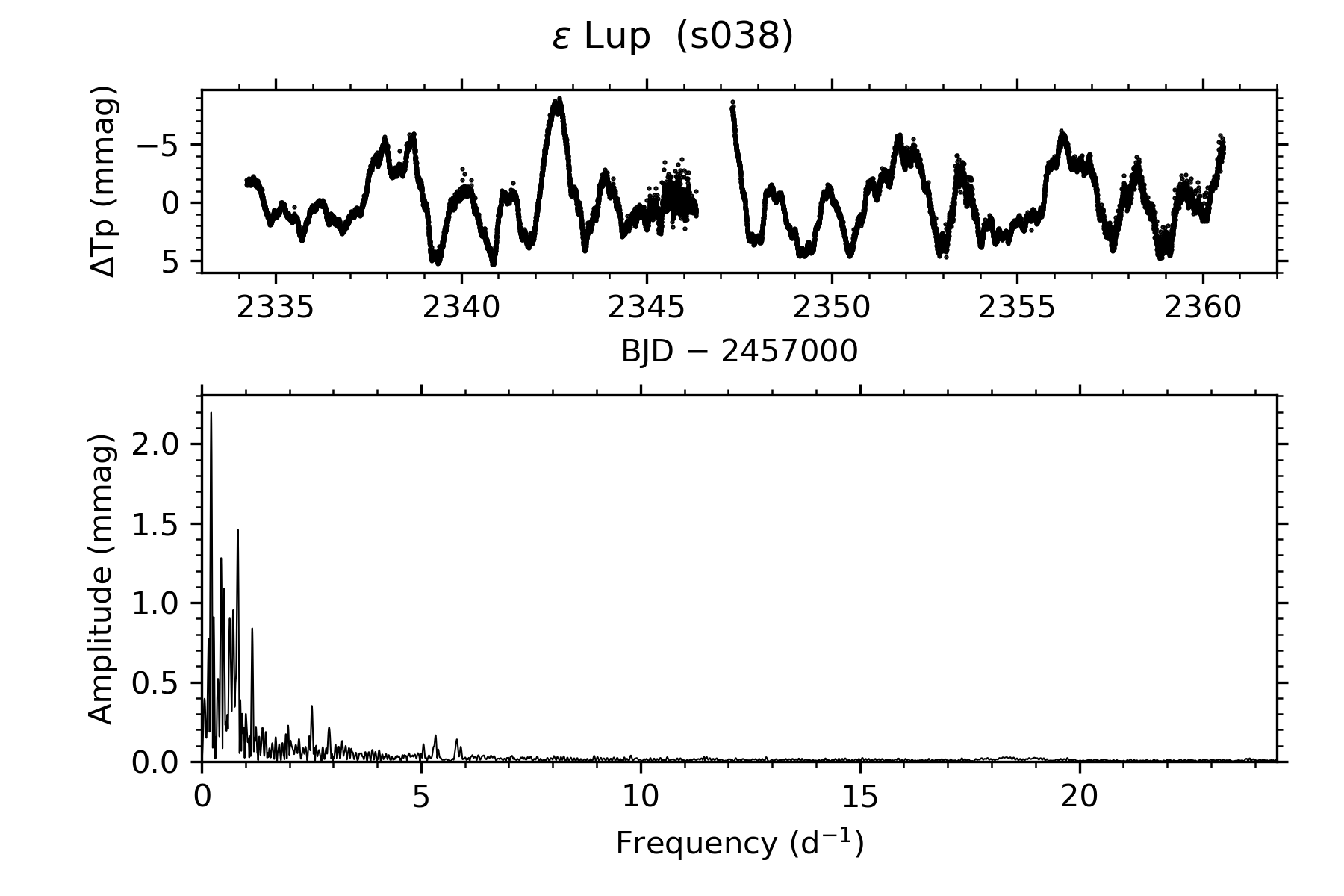}
\includegraphics[width=0.45\textwidth]{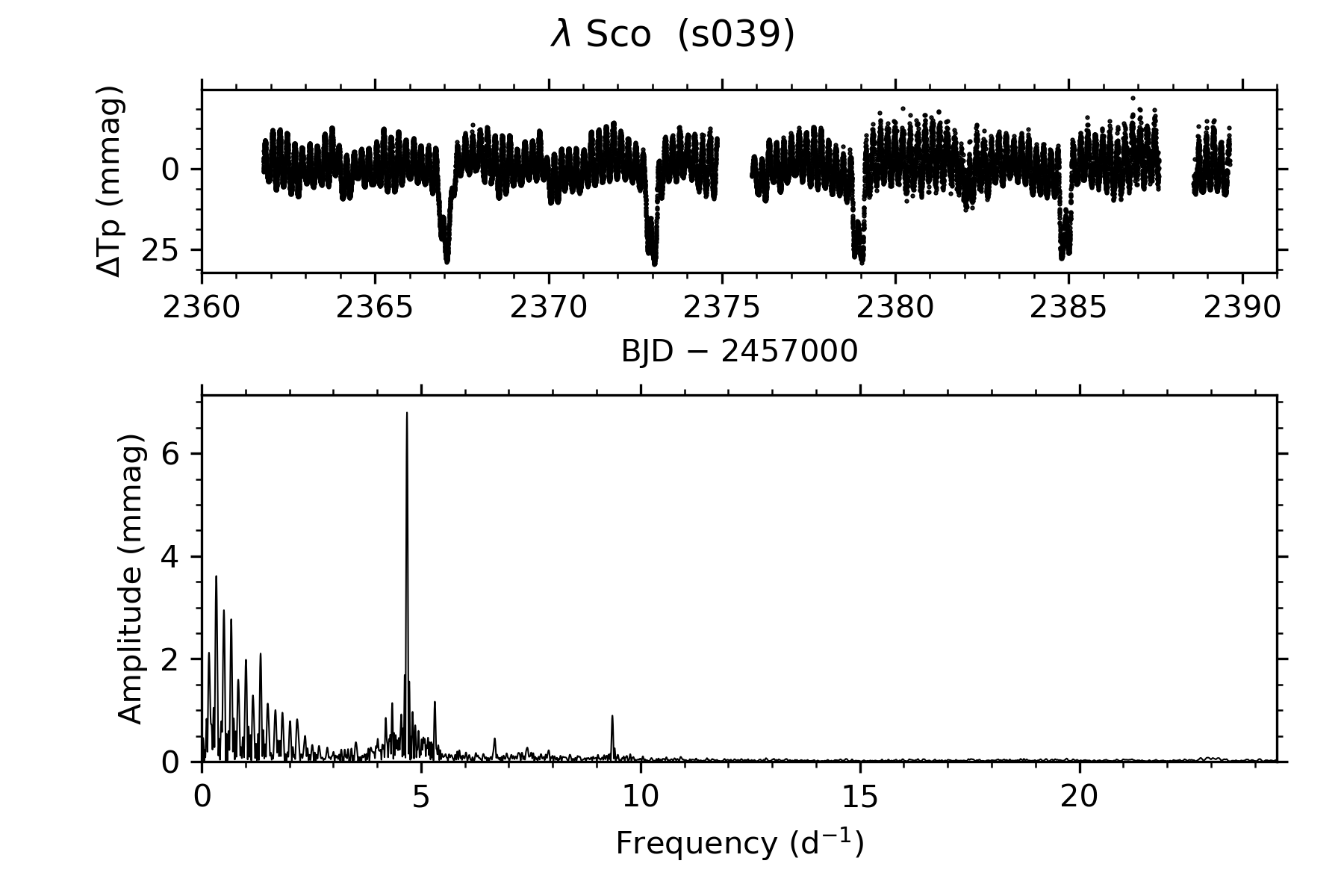}
\includegraphics[width=0.45\textwidth]{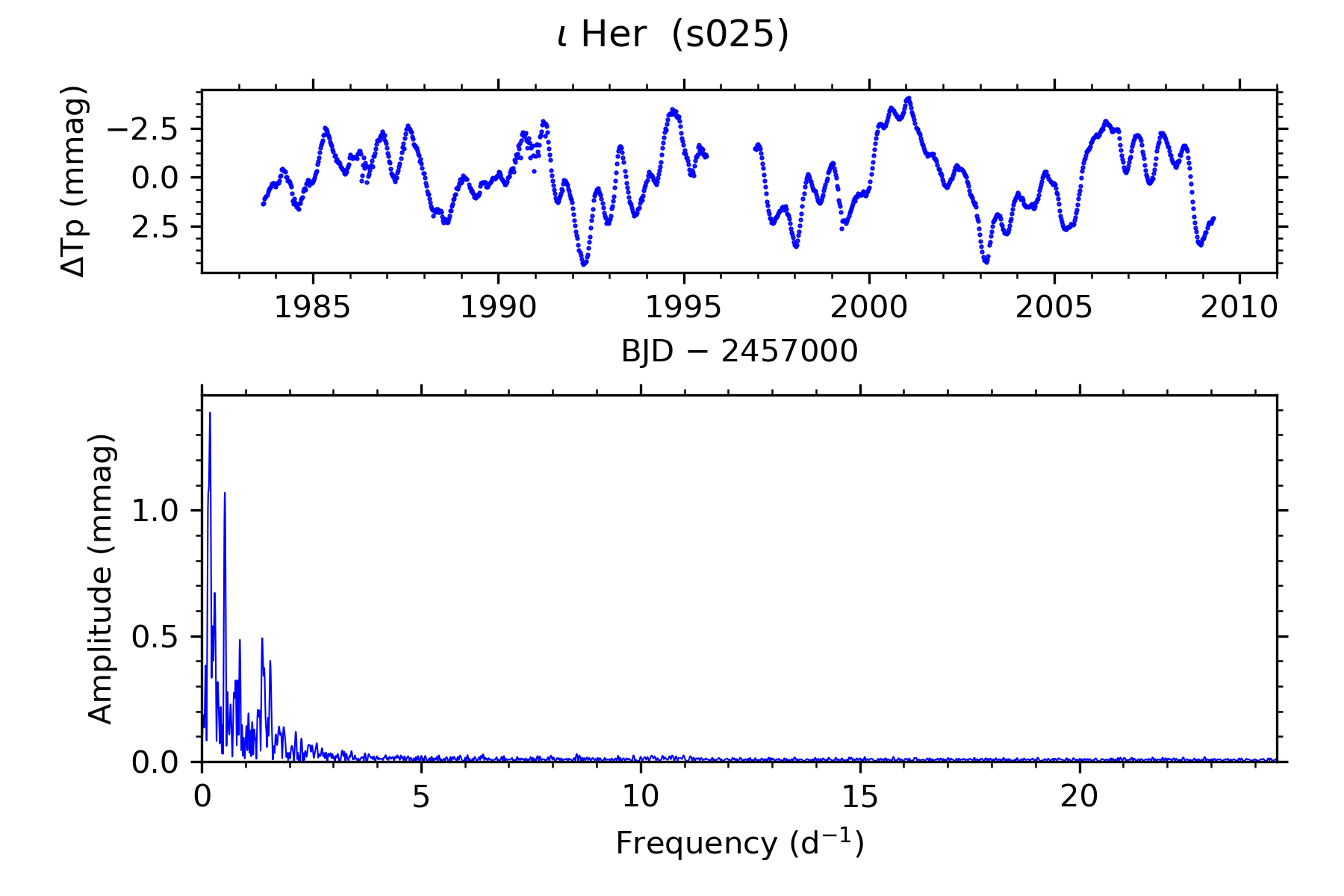}
\includegraphics[width=0.45\textwidth]{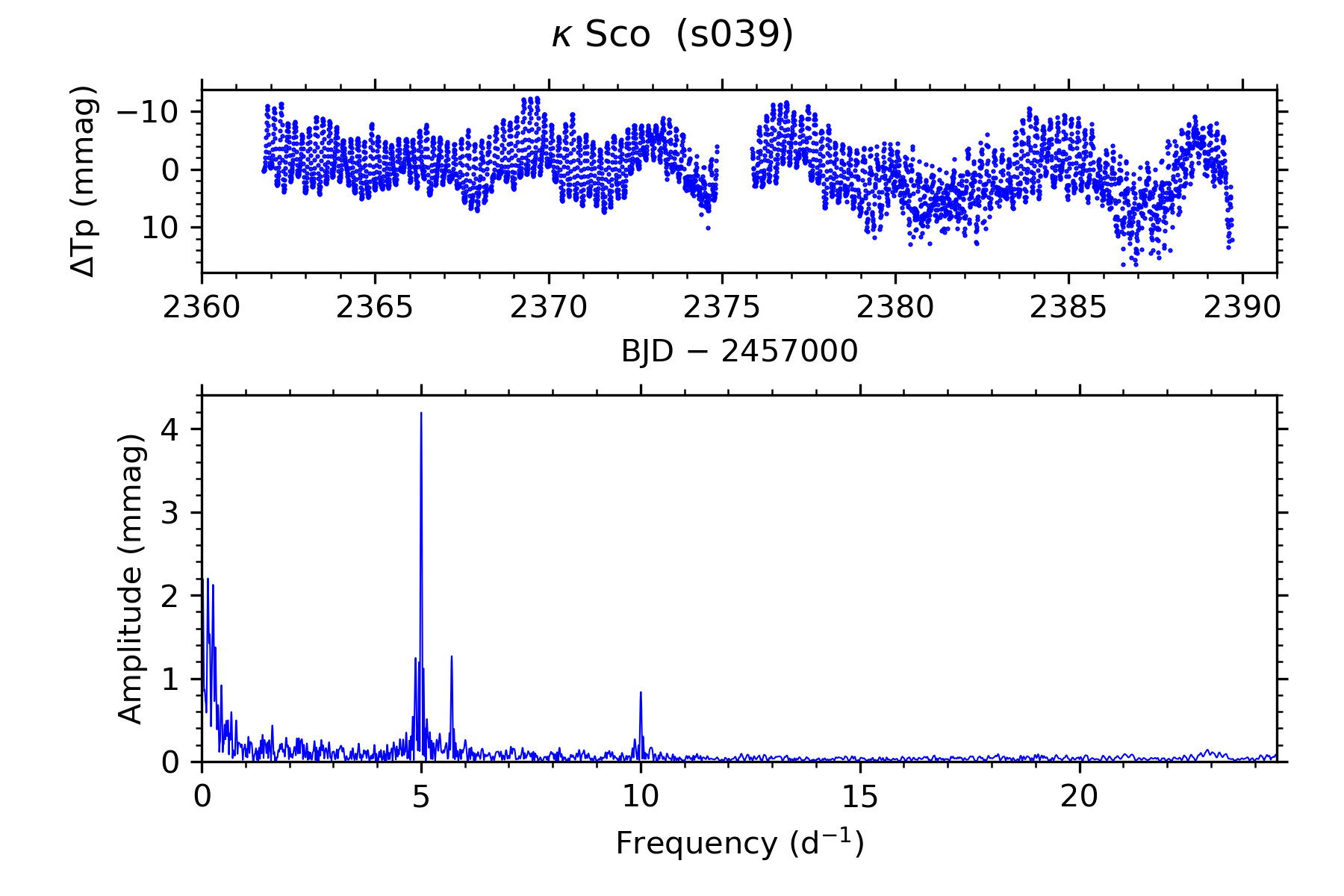}
\includegraphics[width=0.45\textwidth]{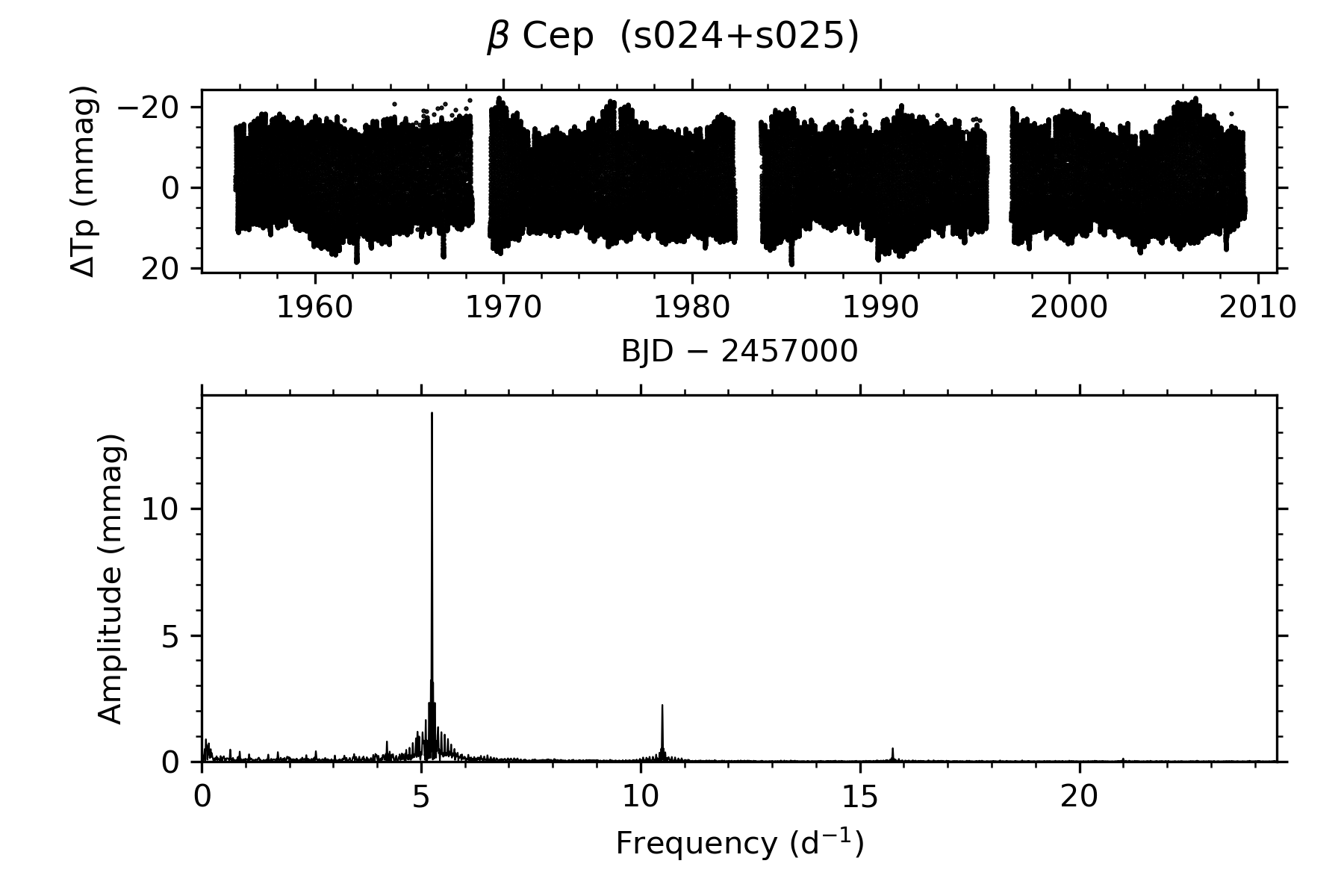}
\caption{TESS light curves and amplitude spectra for priority~1A CubeSpec mission targets. Those shown in black are 2-min cadence PDC-SAP light curves from the SPOC pipeline. Those in blue are light curves extracted using our own simple aperture photometry tools (see text for details).}
\label{figure: priority1_2}
\end{figure*}

The TESS light curves and amplitude spectra of priority~1A targets of the CubeSpec mission are provided in Figs~\ref{figure: priority1_1} and \ref{figure: priority1_2}.

\end{appendix}


\end{document}